\documentclass[a4paper,fleqn,final]{cas-dc}

\usepackage[numbers]{natbib}

\usepackage{subcaption}
\usepackage{commath}
\usepackage{xfrac}
\usepackage{bm}
\usepackage{csquotes}
\usepackage{IEEEtrantools}

\newcommand*\mean[1]{\overline{#1}}
\DeclareMathOperator*{\expected}{\mathbb{E}}
\DeclareMathOperator*{\argmax}{argmax}

\DeclareMathOperator{\sinc}{sinc}

\def\tsc#1{\csdef{#1}{\textsc{\lowercase{#1}}\xspace}}
\tsc{WGM}
\tsc{QE}

\begin{document}
\let\WriteBookmarks\relax
\def\floatpagepagefraction{1}
\def\textpagefraction{.001}

\shorttitle{Array Synthesizer Based on \emph{Slepian} Functions}    

\shortauthors{H.SHARKAS}  

\title [mode = title]{A New Array Synthesizer Based on \emph{Slepian} Functions}  

\tnotemark[1] 

\tnotetext[1]{This paper is entirely based on the author's Master’s thesis work for a self-funded MSc degree in communication systems.
This thesis was conducted with \emph{Huwaei} Technologies Sweden AB, and the degree was obtained from Lund University, Sweden.
The original thesis is publicly available on the university's institutional repository for master’s theses \cite{8975347}.} 

%

\author[1]{Hesham Sharkas}[orcid=0000-0001-5359-4744]

\cormark[1]

\fnmark[1]

\ead{khsharkas@stud.etu.ru; sharkas147@yahoo.com}

\ead[url]{www.researchgate.net/profile/Hesham_Sharkas}


\affiliation[1]{organization={Department of laser measurement and navigation systems, Saint Petersburg electrotechnical university ``LETI''},
            state={Saint Petersburg},
            country={Russia}}

\cortext[1]{Corresponding author}

\fntext[1]{PhD candidate in photonics.}


\begin{abstract}
This study introduces a new multi-antenna array synthesizer based on \emph{Slepian} functions.
The synthesizer concentrates beamforming (BF) gain within a spatial region (i.e., an angular sector), optimizing \emph{Shannon} capacity of the targeted region, which is suitable for codebook-based analog BF.
Starting with the mean capacity formula incorporating the effect of BF, \emph{Jensen}’s inequality was used to find upper and lower bounds of the mean capacity.
Then, a novel method was introduced by combining the two bounds into a new approximation of the mean capacity that outperform both bounds.
Finally, the approximation was formulated to a solvable \emph{Slepian} optimization problem that yielded the weights of the synthesizer.
The properties of the synthesizer were listed, including a discussion on how it behaves by changing the width of the targeted region.
The steering method was derived, and simulation results were presented. 
\end{abstract}



\begin{keywords}
Array synthesizer \sep Beamformer \sep Analog Beamforming \sep Codebook \sep Prolate spheroidal wave functions \sep Slepian functions
\end{keywords}

\maketitle



\section{Introduction}\label{sec:introduction}
Beamforming (BF) is processed in the digital or analog domains.
Digital BF requires a dedicated RF-chain per antenna element making it expensive, complex, and power consuming, however, allows better utilization of spatial degrees of freedom.
Analog BF requires analog phase-shifter and/or variable gain controller per antenna element so it is cheaper and less power consuming, however, less utilization of spatial degrees of freedom (only BF gain) \cite{6736750,6979962}.
Analog BF is commonly implemented using a codebook consists of a set of codewords.
A codeword is a vector of predesigned antennas weights that generates a radiation pattern (i.e., beam/beamform) with desirable shape.
A codebook is designed by quantizing angular coverage to discrete directions and incorporating a codeword for each \cite{6979962,5262295,243458284,7845674}.
Array synthesizers are used to design radiation patterns have advantages -but also disadvantages- for each synthesis method.
Discrete Fourier transform (DFT) synthesizer -also know as uniform array- achieves the highest possible peak value of the main lobe for a certain number of antenna elements, however, high peaks at the side lobes.
Binomial synthesizer shows the widest main lobe with smallest peak value, and no side lobes at all.
\emph{Dolph-Chebyshev} synthesizer has the narrowest main lobe for a given side lobe level.
The three synthesizers maximize the \emph{peak} of the main lobe under some criteria.
Practically, in a codebook-based BF the angle of arrival/departure (AoA/AoD) is not necessarily perfectly aligned with the predesigned discrete directions in the codebook and this causes degradation in performance.
This work tackles this problem by introducing a synthesizer that concentrates BF gain within a given angular sector instead of maximizing the value of the \emph{peak} point.
This allows design a codebook by dividing the angular coverage to angular regions instead of discrete directions, and synthesizing codewords for each region.
In order to concentrate the BF gain within an angular sector, \emph{Slepian} functions \cite{6773659,6773515,1454379,6771595} are used.
The remarkable properties of the functions make them attractive in communications and signal processing.
Many works utilized the functions for waveform design \cite{1264343,7461478}, channel estimation \cite{5522061,5656476,1495893}, signals sampling/recovery \cite{1201663,7394614,7080402,DAVENPORT2012438}, and radar \cite{8950413,6210398}.
For digital BF, this work \cite{1628610} investigated the usage of \emph{Discrete Prolate Spheroidal Sequence} (DPSS) in the \emph{generalized sidelobe canceller} (GSC), and \cite{8631750,9008732} employed \emph{Slepian} optimization to reduce sidelobes of spherical acoustic arrays.

\textbf{Organization of the paper.} The rest of this paper is organized as follows:
In Section \ref{Sec:SystemModel}, system model was described and design assumptions were listed.
In Section \ref{Sec:ProblemFormulation}, The design problem is formulated where \emph{Shannon} capacity is used as performance criteria, then upper and lower bounds on the mean capacity are derived and used to derive an approximation of the mean capacity which is used to formulate the optimization problem of the design.
Section \ref{Sec:Solution} solves the problem to derives the synthesizer weights that maximizes the approximation of the mean capacity, and shows that it maximizes the outage capacity as well.
Then some properties of the synthesizer, as well as the steering method were drived.
By the end of this section, implementation of the synthesizer was discussed.
Simulation results were presented in Section \ref{Sec:Simulation}, and the main results were concluded in Section \ref{Sed:Conclusion}.

\textbf{My contribution.} This paper proposes a new pattern synthesizer that maximizes \emph{Shannon} capacity over an angular sector instead of maximizing the peak at a certain direction which needs perfect alignment.
Moreover, the methodology used in derivations has lots of  potential in other areas of communications (e.g., channel modelling), not only beamforming.
\section{System model and design assumptions}\label{Sec:SystemModel}
\subsection{System model}\label{subsec:ULA}
\begin{figure}[!t]
\centering
\includegraphics[width=2.8in]{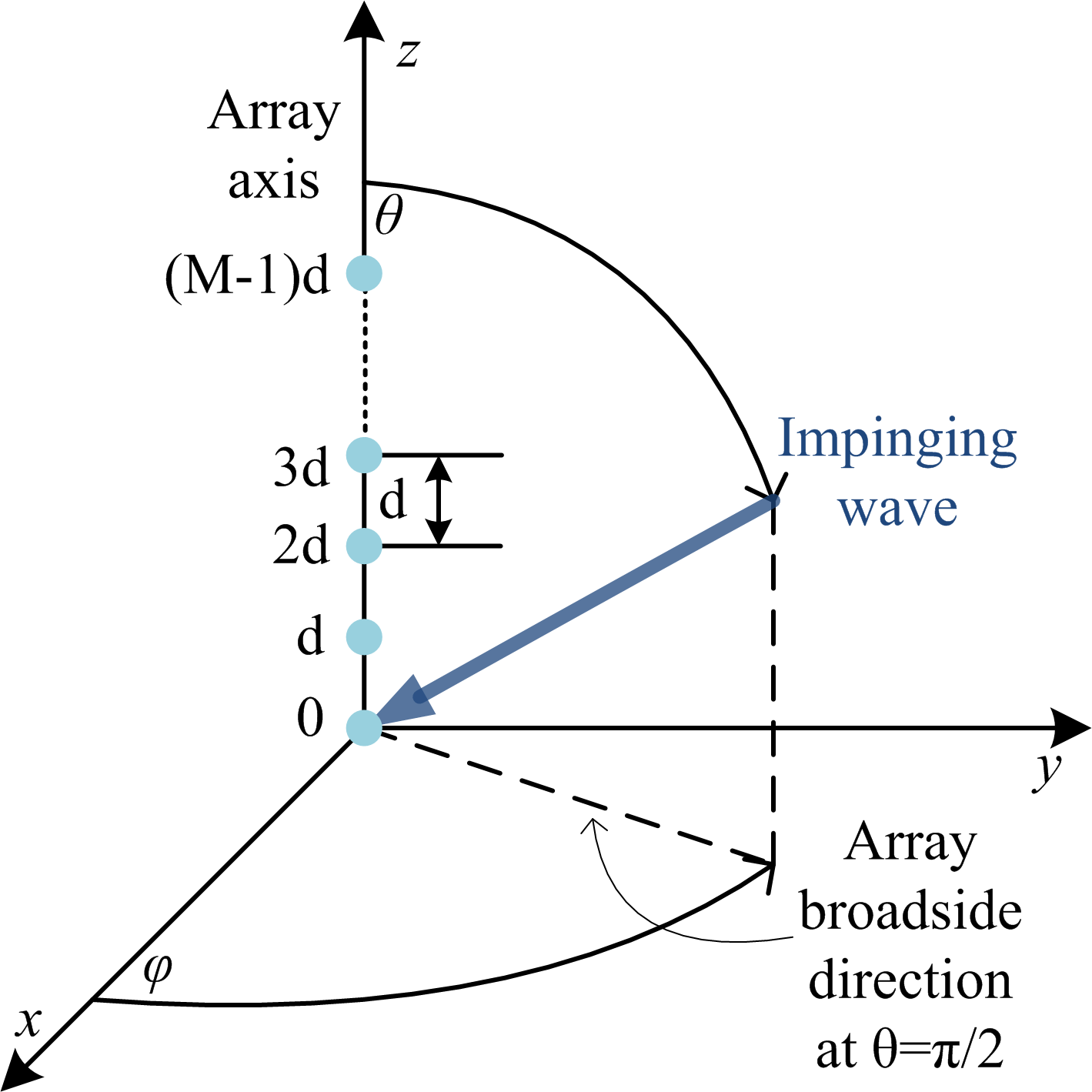}
\caption{A ULA placed along the \emph{z-axis}.}
\label{fig:ULA}
\end{figure}

\begin{figure}
\centering
\begin{subfigure}[t]{1.6in}
\centering
\includegraphics[width=1.5in]{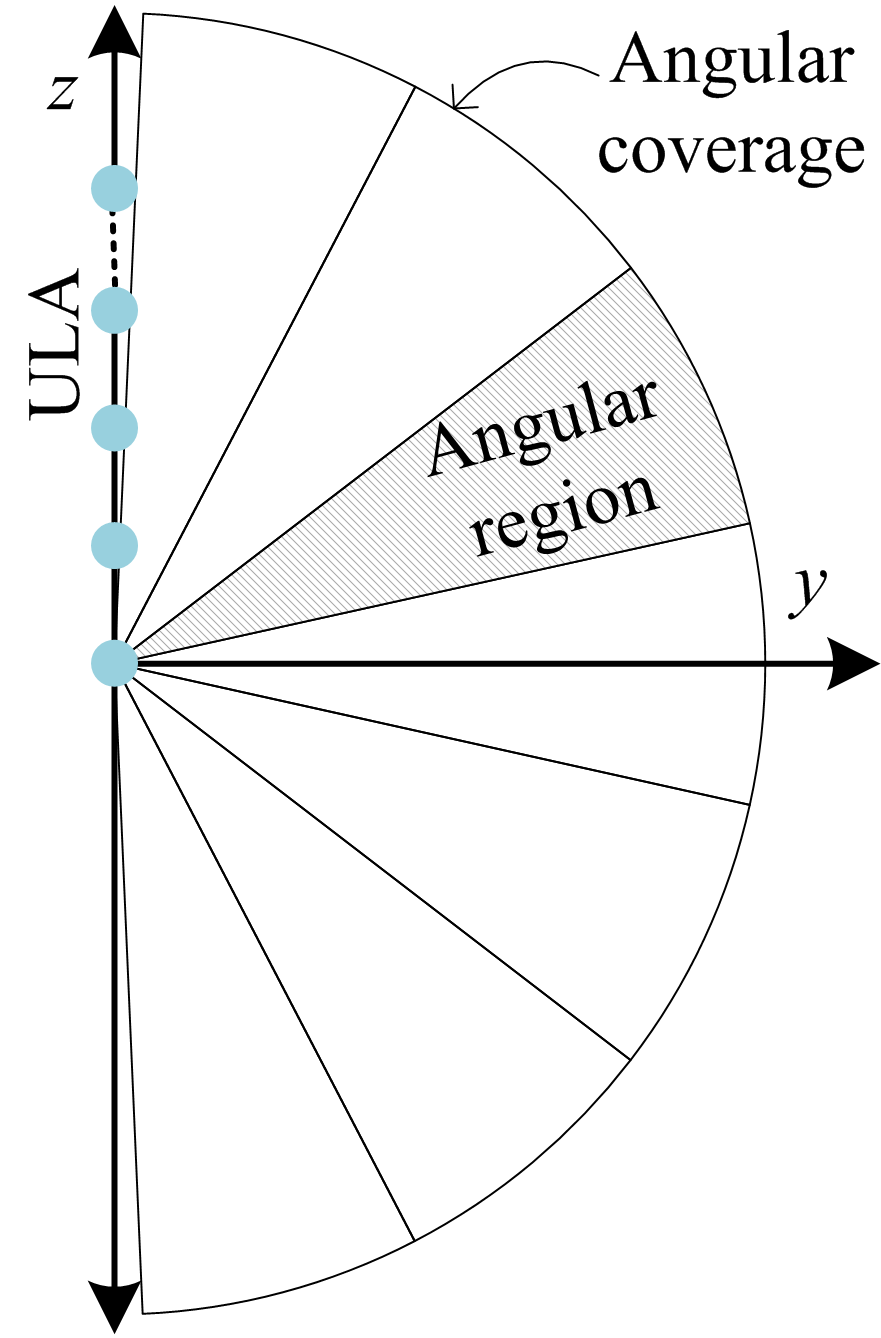}
\caption{Coverage divided to regions.}
\label{fig:AngularCoverage}
\end{subfigure}
\begin{subfigure}[t]{1.6in}
\centering
\includegraphics[width=1.5in]{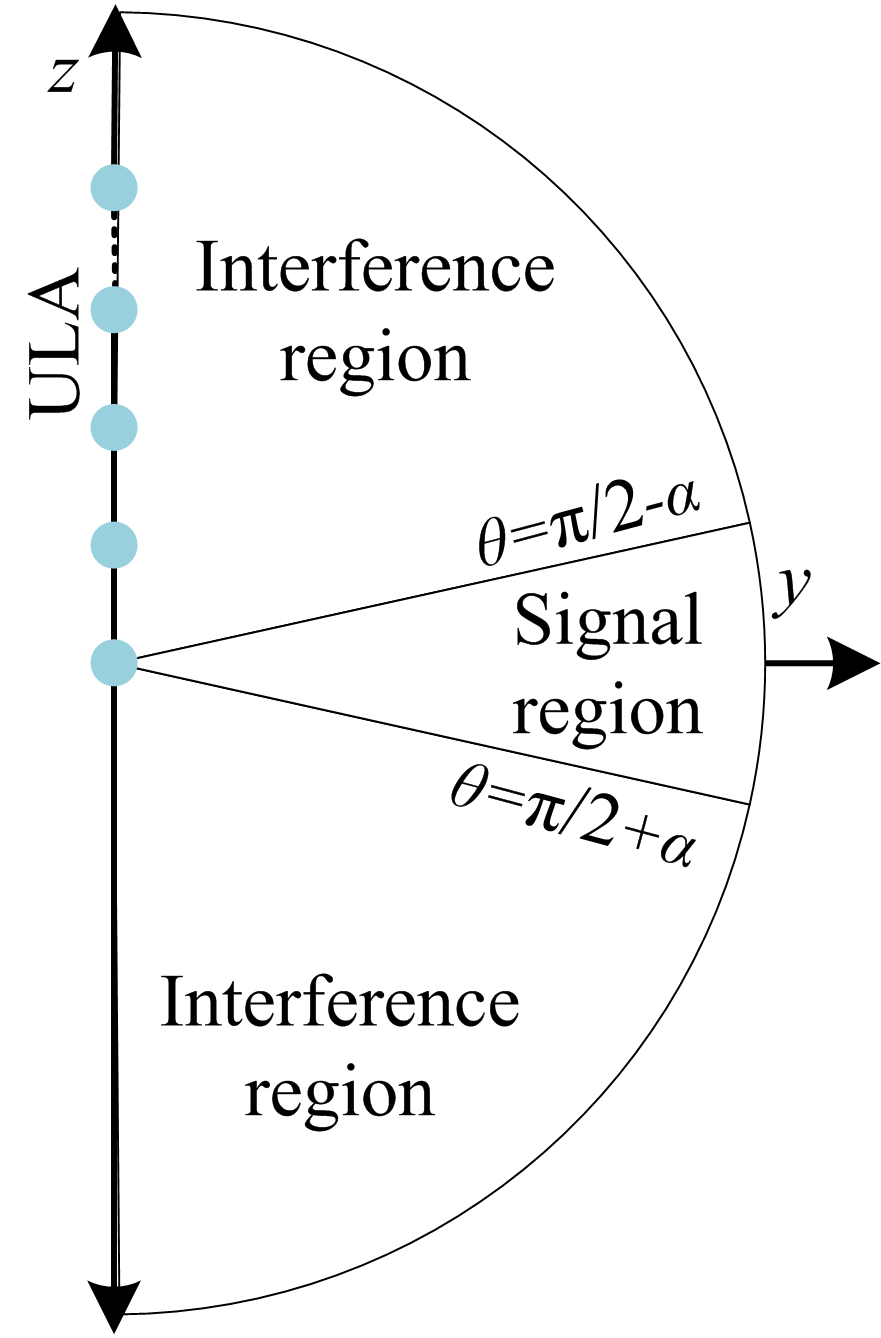}
\caption{Signal/interference regions.}
\label{fig:SignalRegion}
\end{subfigure}
\caption{The angular coverage divided into angular regions instead of discrete directions.}
\label{fig:AngularSpace}
\end{figure}

Assume a uniform linear array (ULA) of $M$ identical antenna elements with equal separation distances $d$.
Assume mutual coupling between antenna elements is neglected.
The ULA is placed along the \emph{z-axis} in \emph{Cartesian} coordinate with the broadside direction at $\theta = \pi / 2$, as shown in Fig.\ref{fig:ULA}.
Such ULA configuration can form beam(s) steering the \emph{zenith} angle $\theta \in [0, \pi]$.
The angular coverage space  $\mathcal{B}_{\theta} = [0, \pi]$ is divided into angular regions $\mathcal{B}_{p}$, each region is characterized by its center angle $p$, see Fig.\ref{fig:AngularCoverage}.
The beam meant to cover $\mathcal{B}_{\pi/2}$ considers the region $\theta=[\pi /2 -\alpha,\pi/2+\alpha]$ as the signal region, and outside (i.e., all other regions) as interference region, see Fig.\ref{fig:SignalRegion}.
The incoming signal may fall anywhere within the signal region.
Next subsection draws design assumptions based on this scenario.
\subsection{Design assumptions}
Assume an incoming desired signal with received power $P_s$, and $N$ number of interfering signals with received power $P_{I,n}$  each, where $n = 1, 2, 3, \dots, N$.
The desired signal has AoA within signal region $\theta_0\in \mathcal{B}_{\pi/2}$, and the interfering signals have AoA outside signal region within the interference region $\theta_n\notin \mathcal{B}_{\pi/2}$, where $n = 1, 2, 3, \dots, N$.
All signals are amplified by the directivity gain based on their AoA.
Assume $\theta_0$ is a random variable with a uniform distribution over the signal region, and $\theta_n$ are random variables with uniform distributions over the interference region,
\begin{eqnarray}
f(\theta_0) & \in & \mathcal{U} (\pi /2 -\alpha\ ,\ \pi/2+\alpha)						\label{DistributionOfSignal}\\
f(\theta_n) & \in & \mathcal{U} (0\ ,\ \pi /2 -\alpha)\cup \mathcal{U}(\pi/2+\alpha\ ,\ \pi)		\label{DistributionOfInterference}
\end{eqnarray}
where the constant $\alpha\in[0,\pi/2]$, $f(.)$ is the probability density function (PDF), and $\mathcal{U}$ is uniform distribution.
All AoA's are assumed independent, so $\theta_0$, $\theta_1$, ..., $\theta_N$ are assumed independent random variables.

Reference \cite{7845674} defined an ideal -but not achievable- beam as a beam that has a flat gain inside the signal region and zero outside, under energy constraint of normalized Frobenius norm of the weights.
Then it used numerical search methods to find the closest achievable beam in shape to the ideal beam, where closeness is measured by minimizing the mean squared error.
The reason ideal beam is not achievable is that a radiation pattern with a square pulse shape needs an infinite number of antenna elements (bandlimited Fourier series needs infinite sequence).
Our goal here is to design a beam that covers, optimizing spectral efficiency, a single region, then steered versions of the beam can be assigned to all other angular regions.
Next subsection utilizes the spectral efficiency as a function of directivity gain which in turn is a function of AoA, to find an optimized beam.
\section{Problem formulation}\label{Sec:ProblemFormulation}
\subsection{Spectral efficiency as performance criteria}
Based on the given ULA geometry and complying with \cite{38.901}, the phase differences between the elements when an impinging planar wave hits the ULA is expressed by a vector $\mathbf{a}$, where
\begin{equation} \label{eq:steering_vector}
\mathbf{a}(\theta) = 
\begin{bmatrix}
1 & e^{-jkd\cos\theta} & e^{-2jkd\cos\theta} & \cdots & e^{(1 - M)jkd\cos\theta}
\end{bmatrix}^T 								\nonumber
\end{equation}
where $k = 2\pi/\lambda$ is the wave number and $\lambda$ is the wavelength.
The weighting vector $\mathbf{v}$ is written as
\begin{equation}
\mathbf{v} = 
\begin{bmatrix}
v_{0} & v_{1} & v_{2} & \cdots & v_{M-1}
\end{bmatrix} 								\nonumber
\end{equation}
The selection of the weighting coefficients in $\mathbf{v}$ controls the shape of the radiation pattern.
The Array factor $AF$ shows how the weights applied to an array influences the total radiation field, and is given by
\begin{equation}
AF(\theta ; \mathbf{v}) = \sum^{M - 1}_{m=0} v_{m} e^{-mjkd\cos\theta } = \mathbf{v} \mathbf{a}(\theta)		\label{eq:ArrayFactor}
\end{equation}
The array directivity gain $G$ (i.e., radiation pattern) is
\begin{equation}\label{eq:DirectivityGain}
G(\theta ; \mathbf{v}) = \abs{AF(\theta ; \mathbf{v})}^2 = \abs{\mathbf{v} \mathbf{a}(\theta)}^2
\end{equation}
Spectral efficiency or \emph{Shannon} capacity per unit bandwidth -mentioned capacity ($C$) in the rest of this work- with the effect of directivity gain  incorporated, is experessed as
\begin{equation}\label{eq:Capacity}
C = \log_2 \big( 1 +  \frac{P_s G(\theta_0 ; \mathbf{v})}{N_0 + \sum_n P_{I,n} G(\theta_n ; \mathbf{v})} \big)
\end{equation}
The interest here is to study the effect of the directivity gain on capacity so $P_s$, $P_{I,n}$, and $N_0$ are considered constants, whereas $\theta_0$ and $\theta_n$ are random variables, \eqref{DistributionOfSignal} and \eqref{DistributionOfInterference}.
Since random variables are present, mean capacity $\mean{C} = \expected \{C\}$ is used instead of $C$, where $\expected \{.\}$ is the expectation operator
\begin{equation}
\mean{C} = \int_{\theta_n}\dots\int_{\theta_1}\int_{\theta_0} f(\theta_0, \theta_1, \dots, \theta_n)\ C\ d\theta_0 d\theta_1 \dots d\theta_n 			\nonumber
\end{equation}
$\theta_0$, $\theta_1$, $\dots$, $\theta_N$ are independent variables so their joint probability function $f(\theta_0, \theta_1, \dots, \theta_n)$ is the multiplication of individual PDFs of each variable.
\begin{equation}
\mean{C} = \int_{\theta_n}\dots\int_{\theta_1}\int_{\theta_0} f(\theta_0)f(\theta_1), \dots, f(\theta_n)\ C\ d\theta_0 d\theta_1 \dots d\theta_n 					\nonumber
\end{equation}
Evaulation of the above integration is hard, instead, $\mean{C}$ is approximated in the next subsection.
\subsection{Approximation of the mean capacity}
It is more tractable to evaluate the bounds instead of the mean capacity.
Jensen's inequality is used to set upper and lower bound to the mean capacity.
Jensen's inequality states
\begin{eqnarray}
\expected\{g(x)\}	&	\leq	&	g(\expected \{x\})\text{\ \ \ \ \ For concave\ }g(x)	\label{eq:jensen_upper}		\\
\expected\{g(x)\}	&	\geq	&	g(\expected \{x\})\text{\ \ \ \ \ For convex\ }g(x)	\label{eq:jensen_lower}
\end{eqnarray}
From \eqref{eq:Capacity}, the mean capacity is given by
\begin{equation}\label{eq:MeanCapacity}
\mean{C} = \expected \big\{ \log_2 \big( 1 +  \frac{\mathbb{S}}{N_0 + \mathbb{I}} \big) \big\}
\end{equation}
where $\mathbb{S}=P_s G(\theta_0 ; \mathbf{v})$ and $\mathbb{I}=\sum_n P_{I,n} G(\theta_n ; \mathbf{v})$.
\subsubsection{Upper bound of the mean capacity}
In order to derive an upper bound of \eqref{eq:MeanCapacity}, the term summed to 1 inside $\log_2$ is considered a variable $x$.
The directivity gain is non-negative as it is square of the array factor \eqref{eq:DirectivityGain}, besides $P_s$, $P_{I,n}$, and $N_0$ are non-negative, so $x$ is non-negative.
A function $\log_2 (1 + x)$ has second derivative $-1/((1+x)^2\ln2)$ that is negative for non-negative $x$ indicating concavity.
Using Jensen's inequality \eqref{eq:jensen_upper}
\begin{eqnarray}
\mean{C} & \leq & \log_2 \Big(1 + \expected \Big\{    \frac{\mathbb{S}}{N_0 + \mathbb{I}}    \Big\}    \Big)											\nonumber\\
		& = & \log_2 \Big(1 + \expected_{\theta_0} \{\mathbb{S}\}\expected_{\theta_n} \Big\{    \frac{1}{N_0 + \mathbb{I}}    \Big\}    \Big) = \mean{C}_{UB}		\label{eq:mean_capacity_upper_bound}			
\end{eqnarray}
The expectation operators are separated because $\theta_0$ and $\theta_N$ are independent.
\subsubsection{Lower bound of the mean capacity}
To derive a lower bound of \eqref{eq:MeanCapacity}, the term summed to 1 inside $\log_2$ is put in the form $\sfrac{1}{x}$, where $x$ is non-negative as above.
A function $\log_2 (1 + \sfrac{1}{x})$ has a second derivative $(2x+1)/((1+x)^2x^2\ln2)$ which is positive for non-negative $x$ indicating convexity.
Using Jensen's inequality \eqref{eq:jensen_lower}
\begin{eqnarray}
\mean{C}  		& = & \expected \Big\{ \log_2 \Big(1 + \frac{1}{ \frac{N_0 + \mathbb{I}}{\mathbb{S}}}   \Big) \Big\}									\nonumber\\
\mean{C}  		& \geq & \log_2 \Big(1 + \frac{1}{\expected \big\{ \frac{N_0 + \mathbb{I}}{\mathbb{S}}\big\}}   \Big)									\nonumber\\
			& = & \log_2 \Big(1 + \frac{1}{\expected_{\theta_0} \{ \frac{1}{\mathbb{S}}\}  \expected_{\theta_n} \{ N_0 + \mathbb{I} \}  }  \Big)				\nonumber\\
			& = & \log_2 \Big(1 + \frac{\frac{1}{\expected_{\theta_0} \{ \frac{1}{\mathbb{S}}\}}}{\expected_{\theta_n} \{ \mathbb{I} \} + N_0 }  \Big)			\nonumber\\
			& = & \log_2 \Big(\frac{\frac{1}{\expected_{\theta_0} \{ \frac{1}{\mathbb{S}}\}} + \expected_{\theta_n} \{ \mathbb{I} \} + N_0}{\expected_{\theta_n} \{ \mathbb{I} \} + N_0 } \Big) =  \mean{C}_{LB}			\label{eq:mean_capacity_lower_bound}
\end{eqnarray}
The upper and lower bounds given in \eqref{eq:mean_capacity_upper_bound} and \eqref{eq:mean_capacity_lower_bound} respectively can be evaluated by calculating the terms $\expected \{1/x\}$ using \emph{Weierstrass Substitution} in conjugate with trigonometric identities for the uniform distribution case \cite[Appendix B]{8975347} independently of the weighting vector, however the solution is dependent on the number of antenna elements and the final formula is complicated and does not give any insights.
Moreover, for general distributions of the desired and interfering signals, the bounds become analytically intractable.
Therefore, a more simple approximation that also outperform both upper and lower bounds is proposed below.
\subsubsection{A new approximation of the mean capacity $\widetilde{\mean{C}}$}
We start by separating the expectation operators in \eqref{eq:MeanCapacity}
\begin{equation}\label{eq:MeanCapacitySeparated}
\mean{C} = \expected_{\theta_n} \big\{ \expected_{\theta_0} \big\{ \log_2 \big( 1 +  \frac{\mathbb{S}}{N_0 + \mathbb{I}} \big) \big\} \big\}
\end{equation}
apply the operator $\expected_{\theta_0}$ to set a  tentative upper bound of \eqref{eq:MeanCapacitySeparated}
\begin{equation}\label{eq:CapacityTentative}
\mean{C} \leq \expected_{\theta_n} \Big\{    \log_2 \Big(1 +         \frac{         \expected_{\theta_0} \{\mathbb{S}\}}{ N_0 + \mathbb{I}  }    \Big)    \Big\} = \mean{C}_{tentative}
\end{equation}
then apply the operator $\expected_{\theta_n}$ to set a lower bound of the obtained upper bound in \eqref{eq:CapacityTentative}
\begin{eqnarray}
\mean{C}_{tentative}  	& \geq &  \log_2 \Big(1 +  \frac{ \expected_{\theta_0} \{\mathbb{S}\}}{ \expected_{\theta_n}\{ N_0 + \mathbb{I} \} }  \Big) = \widetilde{\mean{C}}						\label{MeanCapacityApproximation}\\
\widetilde{\mean{C}} 	& = & \log_2 \Big(1 +  \frac{         \expected_{\theta_0} \{\mathbb{S}\}}{ \expected_{\theta_n}\{ \mathbb{I} \} + N_0} \Big)									\label{MeanCapacityApproximation1}\\
				& = & \log_2 \Big(\frac{  \expected_{\theta_0} \{\mathbb{S}\} + \expected_{\theta_n}\{ \mathbb{I} \} + N_0 }{ \expected_{\theta_n}\{ \mathbb{I} \} + N_0 }    \Big)			\label{MeanCapacityApproximation2}
\end{eqnarray}
The order of the separated expectation operators in \eqref{eq:MeanCapacitySeparated} does not matter, as it leads to the same results in \eqref{MeanCapacityApproximation1} if reversed.
In other words, lower bounding the upper bound leads to the same result as upper bounding the lower bound.\\
In order to prove that the approximation outperforms both upper and lower bounds, firstly, subtract \eqref{MeanCapacityApproximation2} and \eqref{eq:mean_capacity_lower_bound}
\begin{eqnarray}
\widetilde{\mean{C}} - \mean{C}_{LB} 	& = & \log_2 \Big( \frac{ \expected_{\theta_0} \{\mathbb{S}\} + \expected_{\theta_n}\{ \mathbb{I} \} + N_0}{ \expected_{\theta_n}\{ \mathbb{I} \} + N_0 }. 											\nonumber\\
						&     & \ \ \ \ \ \ \ \ \ \ \frac{\expected_{\theta_n} \{ \mathbb{I} \} + N_0}{\frac{1}{\expected_{\theta_0} \{ \frac{1}{\mathbb{S}}\}} + \expected_{\theta_n} \{ \mathbb{I} \} + N_0}  \Big)							\nonumber\\
						& = & \log_2 \Big(\frac{ \expected_{\theta_0} \{\mathbb{S}\} + \expected_{\theta_n}\{ \mathbb{I} \} + N_0}{\frac{1}{\expected_{\theta_0} \{ \frac{1}{\mathbb{S}}\}} + \expected_{\theta_n} \{ \mathbb{I} \} + N_0}\Big) 	\label{Approximation_Sub_LowerBound}
\end{eqnarray}
A function $g(x)=1/x$ is convex for non-negative $x$, so (\ref{eq:jensen_lower}) holds, hence
\begin{equation}
\expected_{\theta_0}\Big\{\frac{1}{\mathbb{S}}\Big\}\geq \frac{1}{\expected_{\theta_0}\{\mathbb{S}\}}\ \ \ \implies\ \ \ \expected_{\theta_0}\{\mathbb{S}\} \geq \frac{1}{\expected_{\theta_0}\big\{\frac{1}{\mathbb{S}}\big\}}		\nonumber
\end{equation}
This means the numerator inside the $\log_2$ function in \eqref{Approximation_Sub_LowerBound} is larger than or equal the denominator.
So the term inside the $\log_2$ function is larger than or equal 1, hence
\begin{eqnarray}
\widetilde{\mean{C}} - \mean{C}_{LB}  	& \geq & \log_2 (1) = 0 							\nonumber\\
\widetilde{\mean{C}}  				& \geq & \mean{C}_{LB} 							\label{eq:approx_lb}
\end{eqnarray}
secondly, subtract \eqref{eq:mean_capacity_upper_bound} and \eqref{MeanCapacityApproximation}
\begin{equation}
\mean{C}_{UB} - \widetilde{\mean{C}} = \log_2 \Big( \frac{1 + \expected_{\theta_0}\{\mathbb{S}\}\expected_{\theta_n}\Big\{    \frac{1}{N_0 + \mathbb{I}}\Big\}}{1 + \expected_{\theta_0}\{\mathbb{S}\}\frac{1}{\expected_{\theta_n}\{N_0 + \mathbb{I} \} }  }  \Big)	 \label{UpperSubtractApproximation}
\end{equation}
A function $g(x)=1/(x+a)$ is convex for non-negative $x$ and $a$, so (\ref{eq:jensen_lower}) holds, hence
\begin{equation}
\expected_{\theta_n}\Big\{\frac{1}{N_0 + \mathbb{I}}\Big\}\geq \frac{1}{\expected_{\theta_n}\{N_0 + \mathbb{I}\}}			 \nonumber
\end{equation}
This means the numerator inside the $\log_2$ function in \eqref{UpperSubtractApproximation} is larger than or equal the denominator.
So the term inside the $\log_2$ function is larger than or equal 1, and
\begin{eqnarray}
\mean{C}_{UB} - \widetilde{\mean{C}} & \geq & \log_2 (1) = 0										 \nonumber\\
\mean{C}_{UB}				& \geq & \widetilde{\mean{C}}									 \label{eq:approx_ub}
\end{eqnarray}
From \eqref{eq:approx_lb} and \eqref{eq:approx_ub}
\begin{equation}
\mean{C}_{UB} \geq \widetilde{\mean{C}} \geq \mean{C}_{LB}									\label{eq:approx_ub_lb}
\end{equation}
This means the approximation is bounded by the upper and lower bounds same as the true mean capacity, so that it is always closer to the true mean capacity than at least one of the bounds - the worst - if not both.
This proves the approximation outperforms the bounds, and it is used in the next subsection to formulate an optimization problem.
\subsection{Formulation of the optimization problem}
\begin{figure}[!t]
\centering
\includegraphics[width=3.3in]{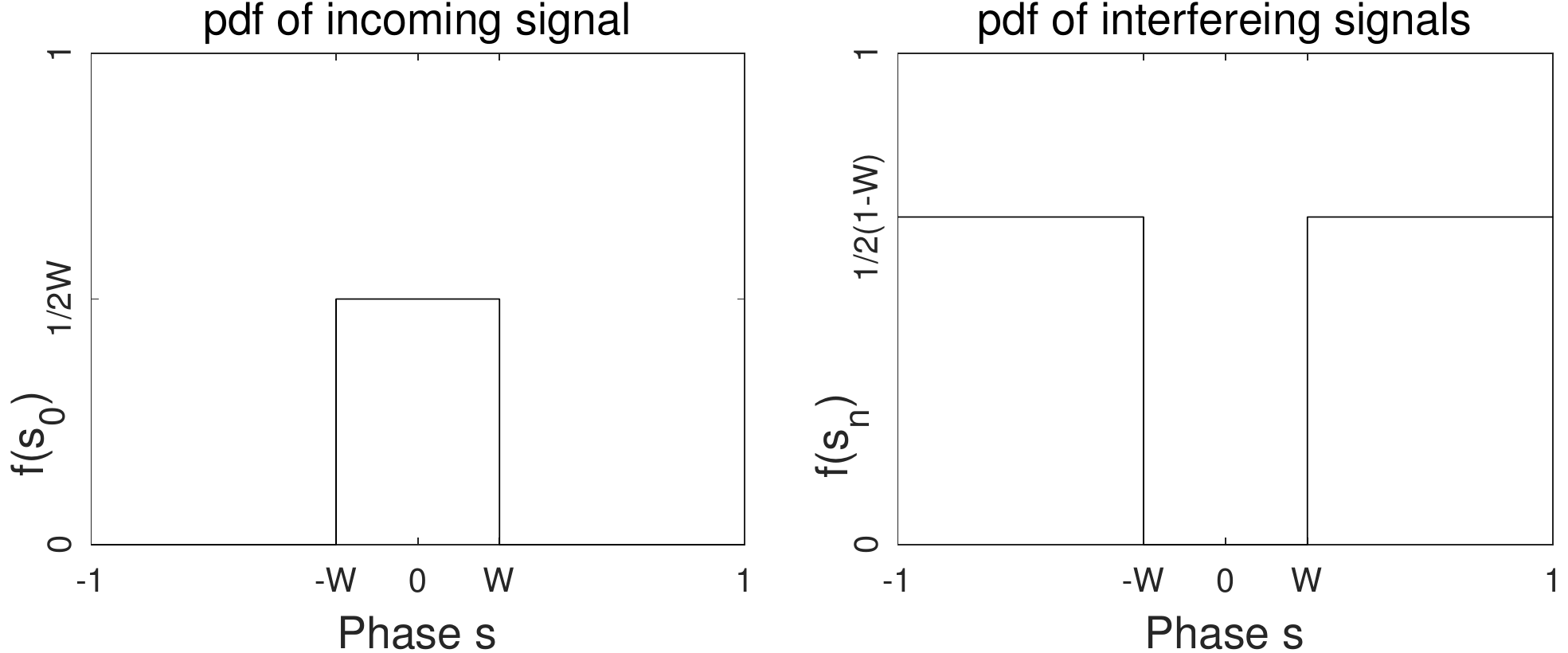}
\caption{PDF of phase for desired and interfereing signals.}
\label{fig:PDFs}
\end{figure}

\begin{figure}[!t]
\centering
\includegraphics[width=3.4in]{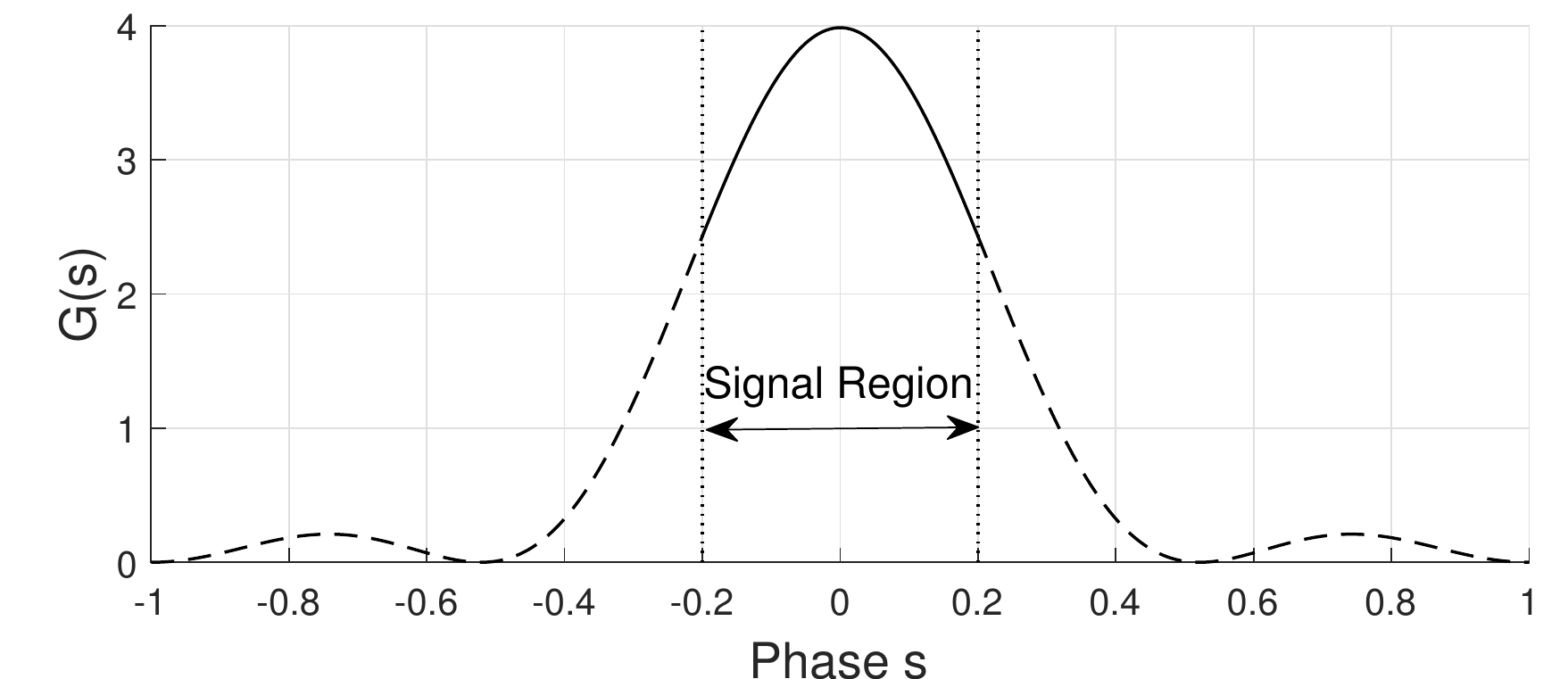}
\caption{Arbitrary directivity gain in the phase domain showing signal/interference sections of the curve ($W=0.2$).}
\label{fig:Arbitrary}
\end{figure}

\begin{figure}[!t]
\centering
\includegraphics[width=3.4in]{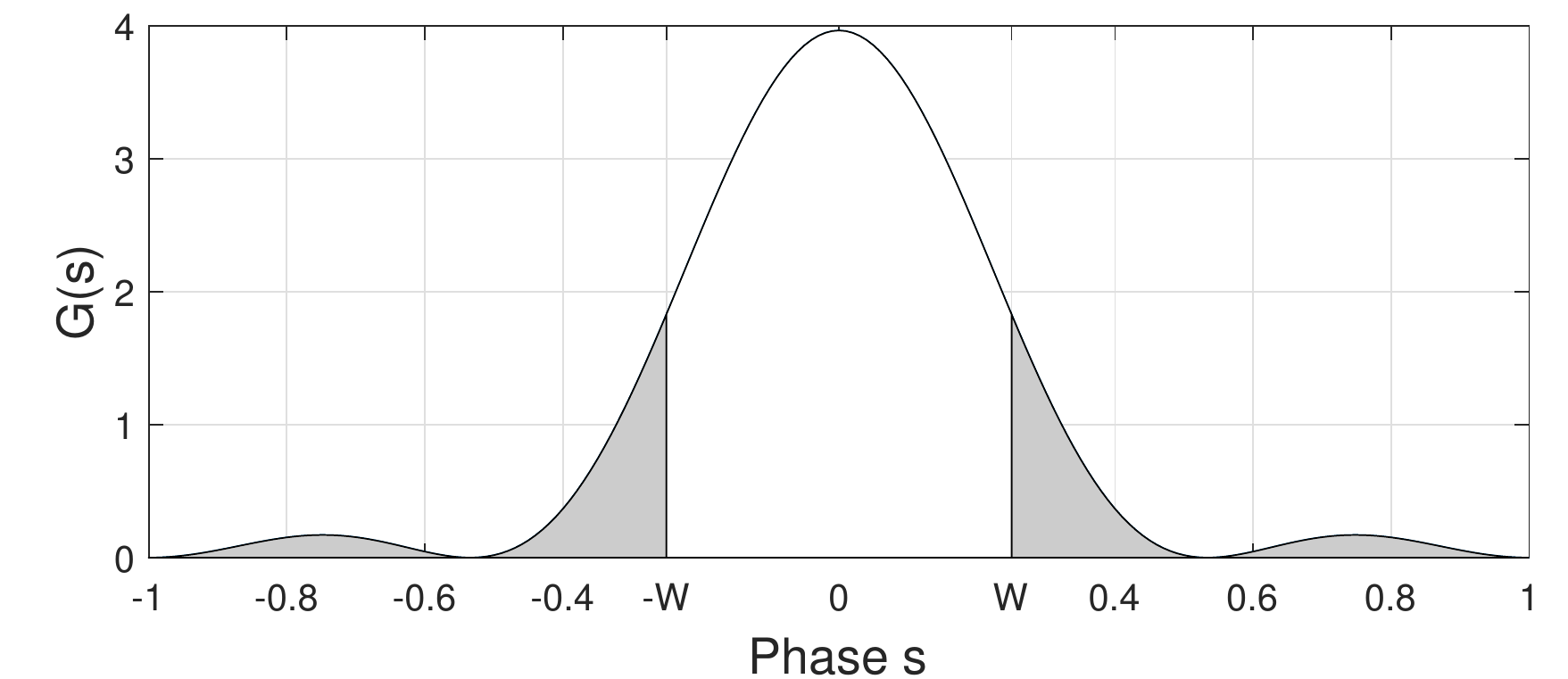}
\caption{Arbitrary directivity gain ($M=4$, $kd=\pi$, $W=0.3$) in the phase domain showing area under signal/interference regions.}
\label{fig:ArbitraryArea}
\end{figure}

From \eqref{eq:DirectivityGain}, directivity gain can be expressed as
\begin{equation}
G(\theta ; \mathbf{v}) = \mathbf{v} \mathbf{a}(\theta) \mathbf{a}^H(\theta) \mathbf{v}^H = \mathbf{v} \mathbf{A}(\theta) \mathbf{v}^H	\nonumber
\end{equation}
where, for $\psi = jkd\cos\theta$,
\begin{equation}
\mathbf{A}(\theta)  =  
\begin{bmatrix}
1					& e^{\psi}			& e^{2\psi}			& \cdots 	& e^{(M-1)\psi}\\
e^{-\psi}				& 1				& e^{\psi}			& \cdots 	& e^{(M-2)\psi}\\
e^{-2\psi}				& e^{-\psi}			& 1				& \cdots 	& e^{(M-3)\psi}\\
\vdots 				& \vdots 			& \vdots 			& \ddots 	& \vdots \\
e^{(1-M)\psi}			& e^{(2-M)\psi}		& e^{(3-M)\psi}		& \cdots 	& 1\\
\end{bmatrix}\nonumber
\end{equation}
Using \eqref{MeanCapacityApproximation1}, and substituting $\mathbb{S}$, $\mathbb{I}$, and $G(\theta ; \mathbf{v})$ replacing $\theta$ by either $\theta_0$ or $\theta_n$ based on the applied expectation.
\begin{IEEEeqnarray}{rcl}
\widetilde{\mean{C}}	& = & \log_2 \Big(1 + \frac{ \expected_{\theta_0} \{P_s  \mathbf{v} \mathbf{A}(\theta_0) \mathbf{v}^H \}}{ N_0 + \expected_{\theta_n}\{ \sum_n P_{I,n}  \mathbf{v} \mathbf{A}(\theta_n) \mathbf{v}^H \} } \Big)				\nonumber\\
				& = & \log_2 \Big(1 + \frac{ P_s  \mathbf{v} \expected_{\theta_0}\{ \mathbf{A}(\theta_0)\} \mathbf{v}^H }{ N_0 + \sum_n P_{I,n}  \mathbf{v} \expected_{\theta_n} \{\mathbf{A}(\theta_n) \} \mathbf{v}^H } \Big)				\label{eq:CapacityExpectation}
\end{IEEEeqnarray}
Evaluating the expectations in the numerator in \eqref{eq:CapacityExpectation}
\begin{equation}
\expected_{\theta_0}\{ \mathbf{A}(\theta_0)\} = \int f(\theta_{0})\mathbf{A}(\theta) d\theta = \frac{1}{2\alpha} \int_{\frac{\pi}{2} -\alpha}^{\frac{\pi}{2} +\alpha} \mathbf{A}(\theta) d\theta		\label{nested_trigonometric_integral}
\end{equation}
Integration of the elements of $\mathbf{A}(\theta)$ (for instance the element $e^{\psi} = \cos(kd\cos\theta) + j\sin(kd\cos\theta)$) does not have a known closed-form because of the nested trigonometric functions.
So a work around is used, which is to change from the angular domain to an approximation called the phase domain.
In the phase domain, $\theta$ is replaced by the variable $s = \cos\theta$.
As $\theta \in [0, \pi]$, then $s \in [-1, 1]$.
The working angular space $\mathcal{B}_{\theta} = [0, \pi]$ is replace by the phase space  $\mathcal{S}_{s} = [-1, 1]$.
The signal region $\mathcal{B}_{\pi/2}$ bounded by $\theta=[\pi /2 -\alpha,\pi/2+\alpha]$ is represent in the phase domain by $\mathcal{S}_{0}$ bounded by $s=[\cos(\pi /2 -\alpha),\cos(\pi/2+\alpha)]=[-W,W]$, where $W\in[0,1]$.
AoA of the desired/interfering signals are replaced by phases $s_0 = \cos\theta_0$, and $s_n = \cos\theta_n$, where $n = 1, 2, 3, \dots, N$.
Since $\theta_0$, $\theta_n$ have uniform distributions, then $s_0$ and $s_n$ are not uniform. However, for convenience  they are kept uniform as below
\begin{eqnarray}
f(s_0) & \in & \mathcal{U}(-W\ ,\ W)						\nonumber\\
f(s_n) & \in & \mathcal{U}(-1\ ,\ -W)\cup \mathcal{U}(W\ ,\ 1)		\nonumber
\end{eqnarray}
Fig.\ref{fig:PDFs} depicts the PDFs of $f(s_0)$ and $f(s_n)$.
An arbitrary directivity gain as a function of $s$ is depicted in Fig.\ref{fig:Arbitrary} showing the signal and inference regions.\\
For $\psi_s = jkds$, $\mathbf{A}(\theta)$ is written in the phase domain as,
\begin{equation}
\mathbf{A}(s)  = 
\begin{bmatrix}
1				& e^{\psi_s}			& e^{2\psi_s}		& \cdots 		& e^{(M-1)\psi_s}\\
e^{-\psi_s}			& 1				& e^{\psi_s}			& \cdots 		& e^{(M-2)\psi_s}\\
e^{-2\psi_s}			& e^{-\psi_s}		& 1				& \cdots 		& e^{(M-3)\psi_s}\\
\vdots 			& \vdots 			& \vdots 			& \ddots 		& \vdots \\
e^{(1-M)\psi_s}		& e^{(2-M)\psi_s}		& e^{(3-M)\psi_s}		& \cdots 		& 1\\
\end{bmatrix}\nonumber
\end{equation}
The approximation in \eqref{eq:CapacityExpectation} in the phase domain is
\begin{IEEEeqnarray}{rcl}
\widetilde{\mean{C}}_{s}	& = & \log_2 \Big(1 + \frac{ \expected_{s_0} \{P_s  \mathbf{v} \mathbf{A}(s_0) \mathbf{v}^H \}}{ N_0 + \expected_{s_n}\{ \sum_n P_{I,n}  \mathbf{v} \mathbf{A}(s_n) \mathbf{v}^H \} } \Big)				\nonumber\\
				& = & \log_2 \Big(1 + \frac{ P_s \expected_{s_0}\{ \mathbf{v} \mathbf{A}(s_0) \mathbf{v}^H \} }{ N_0 + \sum_n P_{I,n} \expected_{s_n} \{ \mathbf{v} \mathbf{A}(s_n) \mathbf{v}^H \} } \Big)				\label{eq:CapacityExpectationPhase}
\end{IEEEeqnarray}
where $\mathbf{v} \mathbf{A}(s_0) \mathbf{v}^H$ is the section of the directivity gain inside the signal region (the solid line of the curve in Fig.\ref{fig:Arbitrary}) and $\mathbf{v} \mathbf{A}(s_n) \mathbf{v}^H$ is the section of the directivity gain outside the signal region (the dashed line of the curve in Fig.\ref{fig:Arbitrary}).\\
To evaluating the expectation in the numerator in \eqref{eq:CapacityExpectationPhase}
\begin{eqnarray}
\expected_{s_0}\{ \mathbf{v} \mathbf{A}(s_0) \mathbf{v}^H \} 	& = & \mathbf{v} \int f(s_0)\mathbf{A}(s) ds \mathbf{v}^H 											\nonumber\\
 										& = & \frac{1}{2W} \mathbf{v} \big( \int_{-W}^W \mathbf{A}(s) ds \big) \mathbf{v}^H							\label{eq:IntegrationPhase}
\end{eqnarray}
Unlike $\mathbf{A}(\theta)$, integration of the elements of $\mathbf{A}(s)$ (for instance $e^{jkds} = \cos(kds) + j\sin(kds)$ - have a closed-form.
Integrating $\cos(kds) + j\sin(kds)$ over symmetric bounds around zero ($\pm W$) results that the imaginary part vanishes as $\sin$ is an odd function, and the real part integrates to $2\sin(kdW)/kd$.
However, one must note:
\begin{enumerate}
\item Integral across $s$ should not be considered \emph{integration by substitution}, because the latter requires change of:
\begin{enumerate}
\item $d\theta$ by ($-ds/\sin\theta$), as																	\label{FirstPoint}
\begin{equation}s = \cos\theta \ \implies\ \frac{ds}{d\theta} = -\sin\theta \ \implies\ d\theta = \frac{-ds}{\sin\theta}\nonumber\end{equation}
\item Integral bounds: For example if the integral is bounded by $[a,b]$, then the bounds should be changed to $[\cos a,\cos b ]$.				\label{SecondPoint}
\end{enumerate}
Although point \ref{SecondPoint} is satisfied, point \ref{FirstPoint} is not, as the appearance of $\theta$ makes the integral still unsolvable, so $ds$ is just used instead of $-ds/\sin\theta$.
Although this is not the same integration problem as in the angular domain, it is still a good approximation \cite{8975347}.
\item\label{numerical_integration_angular_domain} As stated before, the integral in the angular domain has no closed-form.
However, it is definite so it can be evaluated numerically and saved as a look-up table.
This work proceeds in the phase domain to be able to gain insights from closed-form expressions.
The insight of interest is how the matrix $\mathbf{A}(W)$ responds to change of the integral bounds $\pm W$, and how to steer the beam.
\end{enumerate}
The integral in \eqref{eq:IntegrationPhase} is integrable to the matrix $\mathbf{A}(W)$ where,
\begin{IEEEeqnarray}{rcl}
\mathbf{A}(W)_{m,n} & = &
     \begin{cases}
       \frac{2\sin((m-n)kdW)}{(m-n)kd} &, m \neq n 			\\
       2W &, m = n,\ \ \ \ m,n=1, 2,\dots, M				\\
     \end{cases}								\nonumber\\
& = & 2W\sinc((m-n)kdW)							\nonumber
\end{IEEEeqnarray}
where $\sinc(x) = \frac{\sin(x)}{x}$ for $x \neq 0$, and $\sinc(0)=1$.
Matrix $\mathbf{A}(W)$ is real \emph{symmetric Toeplitz} (i.e., diagonal-constant), which necessairly means it is \emph{centrosymmetric} and \emph{persymmetric}.
The determinant of \emph{Centrosymmetric} matrices is discussed in \cite[pp.183, \S 137]{muir2003treatise} and of \emph{persymmetric} in \cite[pp.187, \S 145]{muir2003treatise}.\\
\emph{Lemma 1}: $\mathbf{A}(W)$ is \emph{positive semi-definite} within $1\geq W \geq 0$ and \emph{positive definite} within $1\geq W > 0$, see Appendix \ref{Positive definiteness}.\\
I use the latter throughout this work and consider investigating the case of $W=0$ separately in Section \ref{ScanW}.
Initially, I consider the case $d=\lambda/2$ and hence $kd=\pi$, and $\norm{\mathbf{v}}_2^2 = 1$.
In Section \ref{section:subwavelength} cases of $kd\neq \pi$ are discussed.\\
\emph{Lemma 2}: The area under $G(s;\mathbf{v})$ is constant \eqref{eq:B2} across $\mathcal{S}_{s}$
\begin{equation}
\int_{\mathcal{S}_{s}} G(s ; \mathbf{v})ds  =  \int_{-1}^{1} \abs{\mathbf{v} \mathbf{a}(s)}^2 ds = 2 \norm{\mathbf{v}}_2^2 = 2	 	\nonumber
\end{equation}
where, $\mathbf{a}(s)=\begin{bmatrix}1 & e^{-jkds} & e^{-2jkds} & \cdots & e^{(1 - M)jkds}\end{bmatrix}^T$.
This means regardless of the shape of the directivity gain, the area under the curve in Fig.\ref{fig:Arbitrary} is preserved to a constant value.
The area bounded by $\pm W$ (the white area in Fig.\ref{fig:ArbitraryArea}) equals
\begin{equation}
\text{Area inside}[-W,W] = \mathbf{v} \mathbf{A}(W) \mathbf{v}^H   												\nonumber
\end{equation}
and the expectation in the numerator in \eqref{eq:CapacityExpectationPhase} equals
\begin{equation}
\expected_{s_0}\{ \mathbf{v} \mathbf{A}(s_0) \mathbf{v}^H\} = \frac{1}{2W} \mathbf{v} \mathbf{A}(W) \mathbf{v}^H				\label{eq:ExpectationSignalRegion}
\end{equation}
the area outside $\pm W$ (the grey area in Fig.\ref{fig:ArbitraryArea}) equals the white area subtracted from the constant total area
\begin{eqnarray}
\text{Area outside}[-W,W] & = & 2 - \mathbf{v} \mathbf{A}(W) \mathbf{v}^H   						\nonumber \\
& = & \mathbf{v} 2\mathbf{I}_{M} \mathbf{v}^H - \mathbf{v} \mathbf{A}(W) \mathbf{v}^H   				\nonumber \\
& = &\mathbf{v}     \big (           2\mathbf{I}_{M} -      \mathbf{A}(W)               \big )       \mathbf{v}^H			\nonumber
\end{eqnarray}
where $\mathbf{I}_{M}$ is the $M\times M$ \emph{identity matrix}.
Then the expectation in the denominator in \eqref{eq:CapacityExpectationPhase} equals
\begin{equation}
\expected_{s_n}\{ \mathbf{v} \mathbf{A}(s_n) \mathbf{v}^H \} = \frac{1}{2(1-W)} \mathbf{v} \big( 2\mathbf{I}_{M} - \mathbf{A}(W) \big) \mathbf{v}^H			\label{eq:ExpectationInterferenceRegion}
\end{equation}
then \eqref{eq:CapacityExpectationPhase} becomes
\begin{equation}
\widetilde{\mean{C}}_{s} =  \log_2 \Big(1 + \frac{  \frac{1}{2W}P_s  \mathbf{v} \mathbf{A}(W) \mathbf{v}^H }{ N_0 +  \frac{1}{2(1-W)}  \mathbf{v} ( 2\mathbf{I}_{M} -      \mathbf{A}(W)	) \mathbf{v}^H \sum_n P_{I,n} } \Big)			\nonumber
\end{equation}
To maximize the equation above, it is required to find $\mathbf{v}$ that maximizes the area inside the signal region (numerator), at the same time minimizes the area of the interference region (denominator).
This formulates an optimization problem
\begin{equation}
\mathbf{v}^{opt} = \argmax_{\mathbf{v}:\ \norm{\mathbf{v}}_2^2 = 1}   \frac{\mathbf{v} \mathbf{A}(W) \mathbf{v}^H}{\mathbf{v} \big( 2\mathbf{I}_{M} - \mathbf{A}(W) \big) \mathbf{v}^H} 								\label{eq:optimization_problem}
\end{equation}
\section{Solution of the optimization problem}\label{Sec:Solution}
\emph{Discrete prolate spheroidal wave functions} (DPSWF) \cite{6771595} which was reported in \cite{SLP} to have benefits in beamforming (BF), are proposed as a solution to \eqref{eq:optimization_problem} that maximizes the energy concentration of the directivity gain for the spatial frequencies bounded by $[-W,W]$.
\subsection{Derivation of the synthesizer weights}\label{subsec:DerivationWeight}
Equation \eqref{eq:optimization_problem} is the \emph{Slepian} optimization in discrete time which can be written as \emph{Rayleigh quotient} \cite{8372635}.
$\mathbf{A}(W)$ is real symmetric so it is always diagonalizable.
\emph{Symmetric eigenvalue decomposition} (SED) is used stating that $\mathbf{A}(W) = \mathbf{X}\bm{\Lambda}\mathbf{X}^{T}$, where $\mathbf{X}$ is orthogonal (i.e., $\mathbf{X}^{T}=\mathbf{X}^{-1}$)
\begin{eqnarray}
\mathbf{v}^{opt} & = & \argmax_{\mathbf{v}:\ \norm{\mathbf{v}}_2^2 = 1}   \frac{\mathbf{v} \mathbf{X}\bm{\Lambda}\mathbf{X}^{T} \mathbf{v}^H}{\mathbf{v} \big ( 2\mathbf{I}_{M} - \mathbf{X}\bm{\Lambda}\mathbf{X}^{T} \big ) \mathbf{v}^H}	\nonumber\\
& = & \argmax_{\mathbf{v}:\ \norm{\mathbf{v}}_2^2 = 1}   \frac{\mathbf{v} \mathbf{X}\bm{\Lambda}\mathbf{X}^{T} \mathbf{v}^H}{\mathbf{v} \mathbf{X} \big ( 2\mathbf{I}_{M} - \bm{\Lambda} \big ) \mathbf{X}^{T} \mathbf{v}^H} 	\nonumber
\end{eqnarray}
let $\mathbf{z} = \mathbf{v} \mathbf{X}$   $\implies$   $\mathbf{z}^H = \mathbf{X}^{T} \mathbf{v}^H$  and    $\mathbf{v} = \mathbf{z}\mathbf{X}^{T}$
\begin{equation}
\mathbf{z}^{opt} = \argmax_{\mathbf{z}:\ \norm{\mathbf{z}\mathbf{X}^{T}}_2^2 = 1}   \frac{\mathbf{z} \bm{\Lambda} \mathbf{z}^H}{\mathbf{z} \big ( 2\mathbf{I}_{M} - \bm{\Lambda} \big )  \mathbf{z}^H} \nonumber
\end{equation}
however, $\norm{\mathbf{z}\mathbf{X}^{T}}_2^2 = \mathbf{z}\mathbf{X}^{T}\mathbf{X}\mathbf{z}^H = \mathbf{z}\mathbf{z}^H = \norm{\mathbf{z}}_2^2$
\begin{equation}
\mathbf{z}^{opt} = \argmax_{\mathbf{z}:\ \norm{\mathbf{z}}_2^2 = 1}   \frac{\mathbf{z} \bm{\Lambda} \mathbf{z}^H}{\mathbf{z} \big ( 2\mathbf{I}_{M} - \bm{\Lambda} \big )  \mathbf{z}^H} 		\label{eq:z_maximize}
\end{equation}
A vector $\mathbf{z}=\begin{bmatrix}1&0&0&\cdots\end{bmatrix}$ maximizes \eqref{eq:z_maximize} by selecting the maximum eigenvalue out of the diagonal matrix $\bm{\Lambda}$. So,
\begin{equation}
\mathbf{v}^{opt} = \mathbf{z}^{opt}\mathbf{X}^{T} = \begin{bmatrix}1&0&0&\cdots\end{bmatrix} \begin{bmatrix}\mathbf{x}_1&\mathbf{x}_2&\cdots&\mathbf{x}_{M}\end{bmatrix}^T = \mathbf{x}_1^T		\nonumber
\end{equation}
Hence, the weighting vector $\mathbf{v}$ that maximizes the area inside the signal region is the transpose of the eigenvector corresponding to the maximum eigenvalue of the matrix $\mathbf{A}(W)$ making the area inside the signal region equals the maximum eigenvalue $\lambda_{max}(\mathbf{A}(W))$, and the maximum value of the optimization problem in (\ref{eq:optimization_problem}) becomes $\lambda_{max}(\mathbf{A}(W))/(2-\lambda_{max}(\mathbf{A}(W)))$.
The vector $\mathbf{x}_1^T$ is the newly introduced synthesizer's weights.
Using these weights, the two expectations in \eqref{eq:ExpectationSignalRegion} and \eqref{eq:ExpectationInterferenceRegion} become
\begin{eqnarray}
\expected_{s_0}\{ \mathbf{v} \mathbf{A}(s_0) \mathbf{v}^H\} & = & \frac{1}{2W} \lambda_{max}( \mathbf{A}(W) )								\nonumber\\
\expected_{s_n}\{ \mathbf{v} \mathbf{A}(s_n) \mathbf{v}^H\} & = & \frac{1}{2(1-W)} \big( 2 - \lambda_{max}(\mathbf{A}(W)) \big) 					\nonumber
\end{eqnarray}
then \eqref{eq:CapacityExpectationPhase} is maximized to
\begin{equation}
\widetilde{\mean{C}}_{s} = \log_2 \Big(1 + \frac{  \frac{1}{2W}P_s  \lambda_{max}( \mathbf{A}(W) ) }{ N_0 +  \frac{1}{2(1-W)} ( 2 - \lambda_{max}(\mathbf{A}(W)) ) \sum_n P_{I,n} } \Big)		\nonumber
\end{equation}
\subsection{Maximization of outage capacity} \label{sec:outage_capacity}
$C_{out,q}$ ($q\%$ outage capacity) is defined as the information rate that is guaranteed for $(100-q)\%$ of the realizations, i.e., $Pr(C\leq C_{out,q})=q\%$ \cite[pp.75]{978-1-107-44741-7}
\begin{equation}
Pr(C \geq C_{out,q}) = 1 - \frac{q}{100} = \hat{q}																							\nonumber
\end{equation}
Plug the approximated capacity \eqref{eq:CapacityExpectationPhase} into the left hand side omitting the expectation operator as this is not the mean
\begin{eqnarray}
Pr \Big(  \log_2 \big(1 + \frac{P_s\mathbf{v}\mathbf{A}(s_0)\mathbf{v}^H}{N_0 +  \sum_{n} P_{I,n} \mathbf{v} \mathbf{A}(s_n) \mathbf{v}^H}\big) \geq C_{out,q} \Big) & = & \hat{q}		\nonumber \\
Pr \Big(   \frac{P_s\mathbf{v}\mathbf{A}(s_0)\mathbf{v}^H}{N_0 + \sum_{n} P_{I,n} \mathbf{v} \mathbf{A}(s_n) \mathbf{v}^H} \geq 2^{C_{out,q}} - 1 \Big) & = & \hat{q}				\nonumber
\end{eqnarray}
Let $\tilde{c} = 2^{C_{out,q}} - 1$, $\mathbb{S}_s =\mathbf{v}\mathbf{A}(s_0)\mathbf{v}^H$, and $\mathbb{I}_{s}=\mathbf{v}\mathbf{A}(s_n)\mathbf{v}^H$.
\begin{eqnarray}
Pr \Big(   P_s\mathbb{S}_s \geq \tilde{c} \big( N_0 + \sum_{n} P_{I,n}\mathbb{I}_s  \big) \Big)	 	 & = & \hat{q}			\nonumber \\
Pr \Big(   P_s\mathbb{S}_s -  \tilde{c} \sum_{n} P_{I,n} \mathbb{I}_s \geq \tilde{c}N_0 \Big)			 & = & \hat{q}			\nonumber
\end{eqnarray}
As long as $C_{out,q}>0$ then $\tilde{c}>0$.
Using Markov-Chebyshev inequality, stating that $Pr(X\geq a) \leq \frac{\expected\{X\}}{a}$ for any $a>0$,
\begin{equation}
\begin{split}
Pr \Big( P_s\mathbb{S}_s - \tilde{c} \sum_{n} P_{I,n} \mathbb{I}_s \geq \tilde{c}N_0 \Big) \leq \frac{\expected \{ P_s\mathbb{S}_s - \tilde{c} \sum_{n} P_{I,n}\mathbb{I}_s \}}{\tilde{c}N_0}		\\
 = \frac{ P_s\expected \{\mathbb{S}_s\} -  \tilde{c} \expected \{\mathbb{I}_s\} \sum_{n} P_{I,n} }{\tilde{c}N_0}
 \end{split}	\nonumber
\end{equation}
The synthesizer maximizes $\expected \{\mathbb{S}_s\}$, and minimizes $\expected \{\mathbb{I}_s\}$, so that it pushes the upper bound of the outage capacity to its maximum, giving more margin to the probability that the capacity goes higher than the outage capacity.
\subsection{Properties of the synthesizer}
\subsubsection{Weights of the synthesizer}
Eigenvectors of a real symmetric matrix are real orthonormal.
Eigenvectors of a centrosymmetric matrix are either \emph{symmetric} or \emph{skew-symmetric} \cite[pp.280, \emph{Theorem 2}]{CANTONI1976275}, and for certain restricted form, eigenvectors corresponding to eigenvalues arranged descendingly are alternately \emph{symmetric} and \emph{skew-symmetric} \cite[pp.283-287, \emph{Theorem 5\&6}]{CANTONI1976275}.
Without giving the proof, the latter theorem is extended to include $\mathbf{A}(W)$ then symmetry of the eigenvector $\mathbf{x}_1$ is described with one of the below;\\
- \emph{Symmetric}: for odd $M$; or even $M$ with $\sinc(kdW) \geq 0$.\\
- \emph{Skew-symmetric}: for even $M$ with $\sinc(kdW) < 0$.\\
In short, weights of the synthesizer are real with unit norm, and their symmetry is as described above.
More accurately, the previous statement describes only the amplitudes of the weights, because phases will be added later to steer the beam.
\subsubsection{Scanning of the parameter $W$}\label{ScanW}
$W$ is adjustable and scans from $0$ to $1$.
Here we study the two extreme cases $W$ equals 0 and 1.
$\mathbf{A}(W=0) = \mathbf{0}_{M \times M}$, eigenvalues of such all-zero matrix (zero rank) are zeros, so for $W = 0$ the eigenvector matrix is undetermined.
However, for $W \to 0$, this approximation ($\sin x \approx x$, for $x \to 0$) is valid.
Hence, $\mathbf{A}(W \to 0) \to 2W\mathbf{1}_{M \times M}$ which is a rank one matrix with one non-zero eigenvalue ($\lambda_{max}=2WM$), and its corresponding eigenvector is $\mathbf{x}_1 = \frac{1}{\sqrt{M}}\mathbf{1}_{M \times 1}$ \cite[Eq.42]{4767985}, which is a DFT beam pointing towards the broadside direction.
This is justified as, for $W$ approaching zero, the width of the signal region converges to a single point then the area is maximized by maximizing the peak.
Since the DFT beam has the maximum possible peak ($=M$), the new beam necessarily converges to it.

For $kd=\pi$, $\mathbf{A}(W=1)_{m,n} =  2 \sinc((m-n)\pi)$ where $m,n=1, 2,\dots, M$, so $\mathbf{A}(W=1) = 2\mathbf{I}_{M}$.
Eigenvalues of such diagonal matrix are all  $2$ and the eigenvector matrix can be any unitary matrix.
Hence, for $W=1$, the eigenvector matrix is undetermined.
However, for $W \to 1$, the new beam converges to the binomial beam.
This is concluded by observing how the zeros of the directivity gain behave by scanning $W$.
The directivity gain is square of the array factor \eqref{eq:DirectivityGain}, so both have the same zeros at the same locations.
The following line is quoted from \cite[pp.1375]{6771595} that describes zeros of the \emph{discrete prolate spheroidal wave function},
\begin{displayquote}
\emph{"The DPSWF $U_k(N,W;f)$ has exactly k zeros in the open interval $-W<f<W$ and exactly $N-1$ zeros in $\sfrac{-1}{2}<f\leq\ \sfrac{1}{2}$."}
\end{displayquote}
The function $U_k(N,W;f)$, where $k=0, 1, \dots, N-1$ represents the array factor generated from $N$ array elements, width $W$, and the eigenvector $k$ as a weighting vector.
Eigenvector $k=0$ corresponds to the maximum eigenvalue, and $k=N-1$ corresponds to the minimum eigenvalue.
This implies that the new array factor has $M-1$ zeros in the interval $s = [-1,1)$, however, it has no zeros in the interval $s = (-W,W)$.
In other words, all zeros of the array factor are always outside the signal region bounded by $\pm W$.
By increasing $W$, the zeros are shifted outwards as follows;
For odd number of zeros, one is located at $s=1$, then half of the remaining converge to $s=-1$, and the other half converge to $s=1$.
For even number of zeros, half of them converge to $s=-1$, and the other half converge to $s=1$.
The binomial array factor has all zeros located at $\theta=0$ and mirrored at $\theta=180$, or in the phase domain located at $s=1$ and mirrored at $s=-1$.
This leads to the conclusion that for $W \to 1$ the array factor converges to the binomial array factor of same $M$.
As the DFTs converge to each other, this implies the sequences (weights) also converge to each other.

\begin{figure}[!t]
\centering
\includegraphics[width=3.4in]{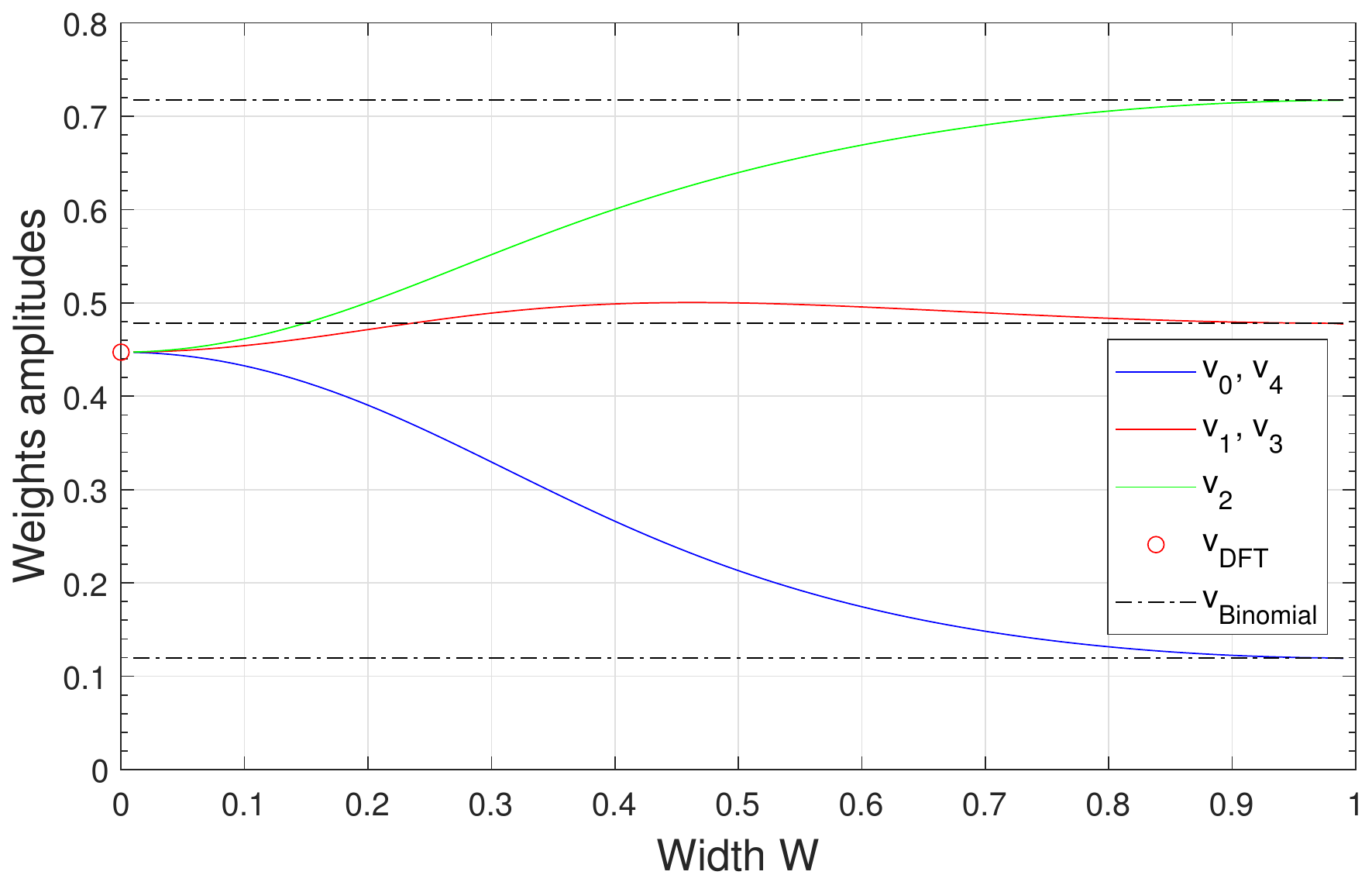}
\caption{The weights at ($M=5$, $kd=\pi$) scan from $W=0$ where they converge to DFT weights, to $W=1$ where they converge to binomial weights.}
\label{fig:weights}
\end{figure}

Fig.\ref{fig:weights} shows graphically the beamwidth $W$ scanning from $0$ to $1$ for a ULA consists of 5 antenna elements.
A DFT synthesizer has equal weights ($v_0=v_1=v_2=v_3=v_4= 1/\sqrt{5} \approx 0.4472$), indicated by a red circle.
From Pascal's triangle, the binomial weight ratios are [1 4 6 4 1], with norm $=\sqrt{70}$.
Weights of the binomial synthesizer are ($v_0=v_4=\frac{1}{\sqrt{70}}\approx 0.1195; v_1=v_3=\frac{4}{\sqrt{70}}\approx 0.4781; v_2= \frac{6}{\sqrt{70}}\approx 0.7171$) depicted by dash-dot lines.
Finally, the weights of the new synthesizer ($kd=\pi$) are drawn versus $W$ and converge to DFT at $W \to 0$ and to binomial at $W \to 1$.
The weights are symmetric, so $v_0$ and $v_4$ coincided, same for  $v_1$ and $v_3$.
\subsection{Steering the beam of the new synthesizer}\label{SteeringSubSection}
\begin{figure}
\centering
\includegraphics[width=3.4in]{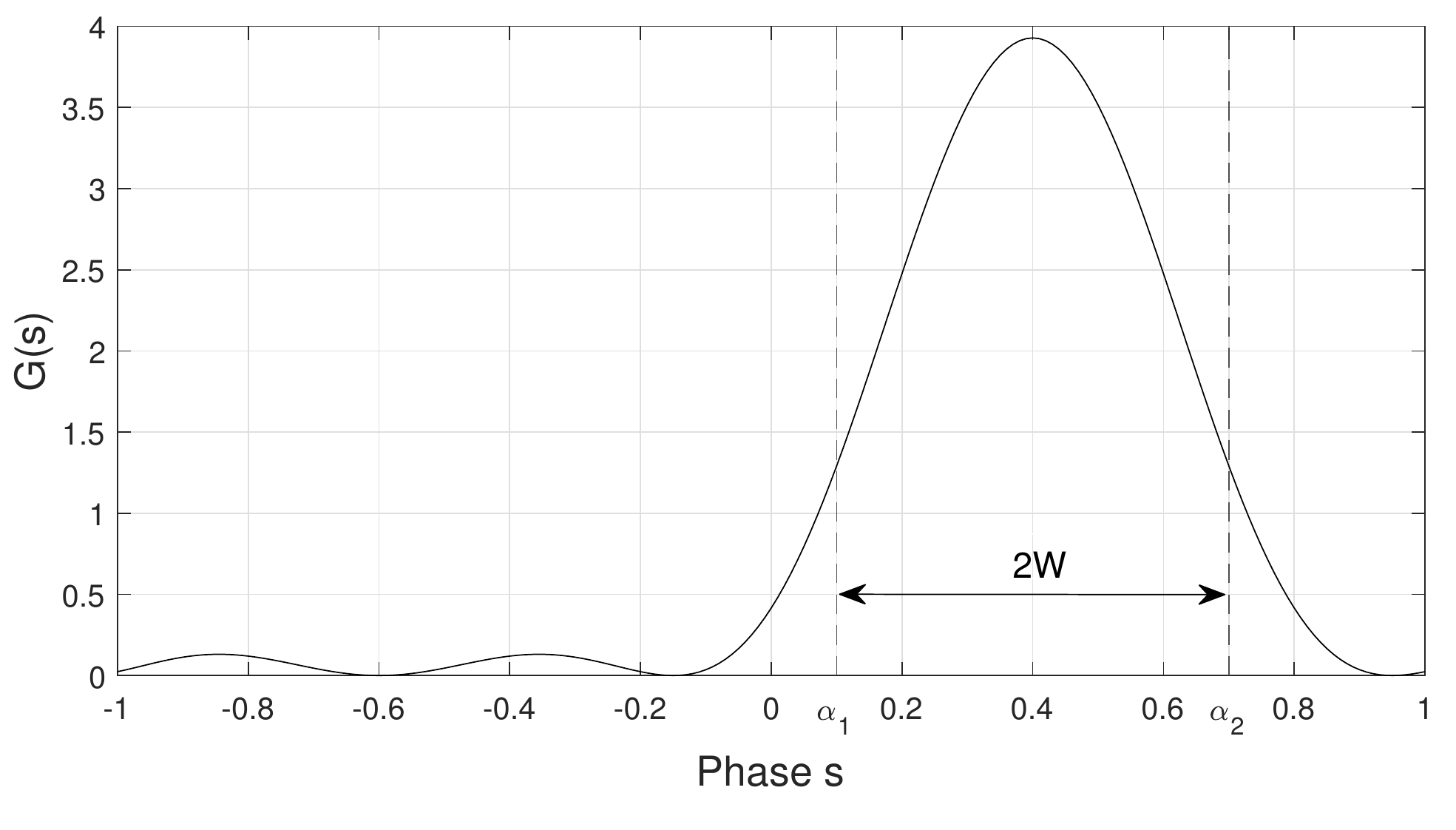}
\caption{A steered version of the new beam ($M=4$, $kd=\pi$, $W=0.3$, $\alpha_1=0.1$, $\alpha_2=0.7$, $s_{s,c}=0.4)$.}
\label{fig:steering}
\end{figure}

So far, the beam is designed for the broadside direction of a ULA.
Steering is simple and does not violate the optimization set in \eqref{eq:optimization_problem}.
Assume a desired signal coming from a steered angular sector with phase $s_s = [\alpha_1, \alpha_2]$ with a uniform distribution $f(s_s) \in \mathcal{U}(\alpha_1 ,\alpha_2)$, where subscript $s$ denotes \emph{steered}.
This changes the integral bounds in \eqref{eq:IntegrationPhase} to
\begin{eqnarray}
\expected_{s_s}\{ \mathbf{v} \mathbf{A}(s_s) \mathbf{v}^H \}	& = & \mathbf{v} \int f(s_s)\mathbf{A}(s)\,ds \mathbf{v}^H 											\nonumber\\
 										& = & \frac{1}{2W} \mathbf{v} \big( \int_{\alpha_1}^{\alpha_2} \mathbf{A}(s) \,ds \big) \mathbf{v}^H					\nonumber
\end{eqnarray}
assuming $\alpha_2 - \alpha_1 = 2W$ as shown in Fig.\ref{fig:steering}.
The integral of the elements of $\mathbf{A}(s)$ across a non-symmetric bounds is
\begin{equation}
\int_{\alpha_1}^{\alpha_2} \mathbf{A}(s)_{m,n} \,ds  =  \int_{\alpha_1}^{\alpha_2} \big[ \cos(ikds)+j\sin(ikds) \big] \,ds		\label{eq:SteeredBounds}
\end{equation}
where $i=n-m$, and $m,n=1,2,3,...,M$ are row and colomn numbers respectively.
Using \emph{integration by substitution} starting with change of variables
\begin{eqnarray}
z & = & s-\alpha_1 - W							\nonumber\\
z|_{s=\alpha_1} & = & \alpha_1 - \alpha_1 - W = -W		\nonumber\\
z|_{s=\alpha_2} & = & 2W + \alpha_1 - \alpha_1 - W = W		\nonumber\\
dz & = & ds									\nonumber
\end{eqnarray}
\begin{equation}
\begin{split}
\int_{\alpha_1}^{\alpha_2} \mathbf{A}(s)_{m,n} \,ds = \int_{-W}^{W} \big[ \cos(ikd(z+\alpha_1 +W))	\\
+ j \sin(ikd(z+\alpha_1 +W)) \big] \,dz
\end{split}	\nonumber
\end{equation}
Let $s_{s,c}=\alpha_1+W$ is the center of the region $s_{s}$.
Substitute with $s_{s,c}$ then use trigonometric identities
\begin{equation}
\begin{split}
= \int_{-W}^{W} \big[ \cos(ikdz)\cos(ikds_{s,c}) - \sin(ikdz)\sin(ikds_{s,c})	\\
+ j \sin(ikdz)\cos(ikds_{s,c}) + j \cos(ikdz)\sin(ikds_{s,c}) \big] \,dz
\end{split}	\nonumber
\end{equation}
second and third terms vanish for symmetric bounds integal
\begin{IEEEeqnarray}{l}
= \frac{2}{ikd}\sin(ikdW) \big[ \cos(ikds_{s,c})+ j\sin(ikds_{s,c}) \big]			\nonumber\\
= \frac{2}{ikd}\sin(ikdW)  e^{jikds_{s,c}} = 2W\sinc(ikdW)  e^{jikds_{s,c}} 		\nonumber
\end{IEEEeqnarray}
Expand the result above to all elements of the matrix $\mathbf{A}(s)$
\begin{equation}
\int_{\alpha_1}^{\alpha_2} \mathbf{A}(s) \,ds  =  \mathbf{A}(W)\circ \mathbf{A}(s=s_{s,c}) 		\label{eq:ShiftSpatialFrequency}
\end{equation}
where $\circ$ is the \emph{Hadamard} product.
$\mathbf{A}(s=s_{s,c})$ is the outer product of the vector $\mathbf{a}(s_{s,c})$ and its Hermitian.
\begin{eqnarray}
\int_{\alpha_1}^{\alpha_2} \mathbf{A}(s) \,ds  & = & [\mathbf{X}\bm{\Lambda}\mathbf{X}^T]\circ [\mathbf{a}(s_{s,c})\mathbf{a}^H(s_{s,c})] 		\nonumber\\
& = & \mathbf{D}_{\mathbf{a}(s_{s,c})} \mathbf{X}\bm{\Lambda}\mathbf{X}^T \mathbf{D}_{\mathbf{a}^*(s_{s,c})} 						\label{eq:HadamardProductDyad}
\end{eqnarray}
where \eqref{eq:HadamardProductDyad} is derived from \cite[pp.479, Lemma 7.5.2.]{horn1994topics}, and a diagonal matrix $\mathbf{D}_\mathbf{u}$ has a vector $\mathbf{u}$ in its main diagonal and zero elsewhere.
As in subsection \ref{subsec:DerivationWeight}, the term $(\mathbf{v}\mathbf{D}_{\mathbf{a}(s_{s,c})} \mathbf{X}\bm{\Lambda}\mathbf{X}^T \mathbf{D}_{\mathbf{a}^*(s_{s,c})}\mathbf{v}^H)$ is maximized by using $\mathbf{z}^{opt} = \begin{bmatrix}1&0&0&\cdots\end{bmatrix}$  where $\mathbf{z}=\mathbf{v}\mathbf{D}_{\mathbf{a}(s_{s,c})} \mathbf{X}$, so
\begin{IEEEeqnarray}{l}
\mathbf{v}^{opt}  = \mathbf{z}^{opt} \mathbf{X}^T \mathbf{D}_{\mathbf{a}(s_{s,c})}^H = \mathbf{x}_1^T	 \mathbf{D}_{\mathbf{a^*}(s_{s,c})} = [\mathbf{x}_1 \circ \mathbf{a^*}(s_{s,c})]^T							\nonumber\\
= \begin{bmatrix}x_{1,1} & x_{1,2}e^{jkds_{s,c}} & x_{1,3}e^{2jkds_{s,c}} & \cdots & x_{1,M}e^{(M\text{-}1)jkds_{s,c}}\end{bmatrix} 																\nonumber
\end{IEEEeqnarray}
which is the real amplitudes of the beam designed for the broadside direction, multiplied \emph{element-wise} by a progressive phase steering vector of phase $s_{s,c}$.
This is restate of \emph{DFT shift theorem}, but instead of shifting the sequence in time it is directly multiplied by the \emph{linear phase term} $\mathbf{a}^*(s_{s,c})$ leading to a shift in spatial frequencies \cite{4767985}.
Spectral magnitude is unaffected by a linear phase term so \eqref{eq:B2} holds consequently \eqref{eq:optimization_problem} is not violated.
Area inside the signal region is the same $\lambda_{max}(\mathbf{A}(W))$.
Fig. \ref{fig:steering} depicts a steered version of the beam.
\subsection{Subwavelength separation distance \lowercase{$\mathbf{d}$}}\label{section:subwavelength}
The array factor \eqref{eq:ArrayFactor} is in the form of \emph{Finite Fourier Series} and is periodic over an interval $s_T=\lambda/d$.
Moreover, the area under the curve $G(s;\mathbf{v})$ within $s_T$ is constant \eqref{eq:B3} and equals $\lambda/d$.
Derivations done so far are for the case ($d=\lambda/2$; $kd=\pi$), where $s_T=[-1,1]$ coincides with the visible region $\mathcal{S}_{s}$.
This section discusses cases ($kd\neq \pi$) whereas $\mathcal{S}_{s}$ is the same.
\subsubsection{\lowercase{$0 < kd \leq \pi$}}
For $0 < d \leq \lambda/2$, then $s_T \in [2,\infty)$.
For some value of $\lambda/d$ in this range, $s_T=[-\lambda/2d,\lambda/2d] > \mathcal{S}_{s}$.
The beamwidth is limited by $\mathcal{S}_s$ so $W \in [0,1]$.
$s_{s,c} \in [-1+W,1-W]$ because beyond this interval the signal region exceeds $\mathcal{S}_{s}$ and disappear in the invisible region.
Taking in consideration the point above, \eqref{eq:optimization_problem} is still a \emph{Rayleigh quotient} but changed to
\begin{IEEEeqnarray}{rcl}
\mathbf{v}^{opt} & = & \argmax_{\mathbf{v}:\ \norm{\mathbf{v}}_2^2 = 1} \frac{\mathbf{v} \mathbf{A}(W) \mathbf{v}^H}{\mathbf{v} \big[ \frac{\lambda}{d}\mathbf{I}_{M} - \underset{I}{\int} \mathbf{A}(s)ds - \underset{II}{\int} \mathbf{A}(s)ds - \mathbf{A}(W) \big] \mathbf{v}^H} 			\nonumber
\end{IEEEeqnarray}
$s_T > \mathcal{S}_s$ so the area under $G(s;\mathbf{v})$ within $\mathcal{S}_s$ is the area within $s_T$ (i.e., $\lambda/d$) subtract two stripes expressed by $I$ and $II$.
The bounds $I$ are $-\lambda/2d - s_{s,c}$ to $-1- s_{s,c}$, and the bounds $II$ are $1 - s_{s,c}$ to $\lambda/2d - s_{s,c}$.
$I$ and $II$ are always outside the signal region, so the area within the stripes are always minimum for any $s_{s,c}$ and $W$.
However, since they are subtracted from the constant $\lambda/2$ this further minimize the denominator in \eqref{eq:optimization_problem}.
\subsubsection{\lowercase{$\pi < kd \leq 2\pi(1-W)$}}
For $\lambda/2<d\leq \lambda(1-W)$, $s_T \in [(1-W)^{-1},2)$.
For some value of $\lambda/d$ in this range, $s_T=[-\lambda/2d,\lambda/2d] < \mathcal{S}_{s}$.
This adds two restrictions:\\
1) The beamwidth is limited by $s_T$ so $W \in [0,\lambda/2d]$.\\
2) Steering phase is limited $s_{s,c} \in [-(\lambda/d)+1+W,(\lambda/d)-1-W]$ because beyond this interval another undesired copy of the steered signal region $s_s$ appears in $\mathcal{S}_{s}$ (grating lobes).\\
Taking the two points in consideration, \eqref{eq:optimization_problem} is still a \emph{Rayleigh quotient} but changed to
\begin{IEEEeqnarray}{rcl}
\mathbf{v}^{opt} & = & \argmax_{\mathbf{v}:\ \norm{\mathbf{v}}_2^2 = 1} \frac{\mathbf{v} \mathbf{A}(W) \mathbf{v}^H}{\mathbf{v} \big[ \frac{\lambda}{d}\mathbf{I}_{M} + \underset{I}{\int} \mathbf{A}(s)ds + \underset{II}{\int} \mathbf{A}(s)ds - \mathbf{A}(W) \big] \mathbf{v}^H} 			\nonumber
\end{IEEEeqnarray}
Since $s_T < \mathcal{S}_s$, the area under $G(s;\mathbf{v})$ within $\mathcal{S}_s$ is the area within $s_T$ (constant and equals $\lambda/d$) plus two stripes expressed by $I$ and $II$.
The bounds $I$ are $-1 - s_{s,c}$ to $-\lambda/2d - s_{s,c}$, and the bounds $II$ are $\lambda/2d - s_{s,c}$ to $1 - s_{s,c}$.
$I$ and $II$ are always outside the signal region, so the area within the stripes are always minimum for any $s_{s,c}$ and $W$.
\subsection{Implementation of the synthesizer}
Assume the singular value decomposition (SVD) of $\mathbf{A}(W) = \mathbf{U} \boldsymbol{\Sigma} \mathbf{V}^{H}$ and the eigen value decomposition (EVD) of $\mathbf{A}(W) = \mathbf{X} \boldsymbol{\Lambda} \mathbf{X}^{H}$.
For a real symmetric \emph{positive-definite} matrix, $\mathbf{X}=\mathbf{U}=\mathbf{V}$ and $\boldsymbol{\Sigma}=\boldsymbol{\Lambda}$.
Therefore, iterative algorithms for both SVD and EVD can be used to obtain the weights.
More customized and faster algorithms for DPSS are proposed in \cite{4767985,330397,WALTER2005432}.
For codebook-based BF, this method is more convenient as the codebook is compiled offline by solving the eigen problem of $\mathbf{A}(W)$ for different values of $W$ as required to obtain the amplitudes.
The phases are multiplied \emph{element-wise}, as described in Section \ref{SteeringSubSection}, to obtain steered versions of the beams.
Moreover, closed-form for the eigenvector $\mathbf{x}_1$ can be derived for small values of $M$ -up to 9- and implemented on a DSP/FPGA for real-time beamwidth control, which is more convenient for beam/target tracking algorithms, particularly in highly time-variant channels.
\section{Simulation results}\label{Sec:Simulation}
\begin{figure}[!t]
\centering
\includegraphics[width=3.34in]{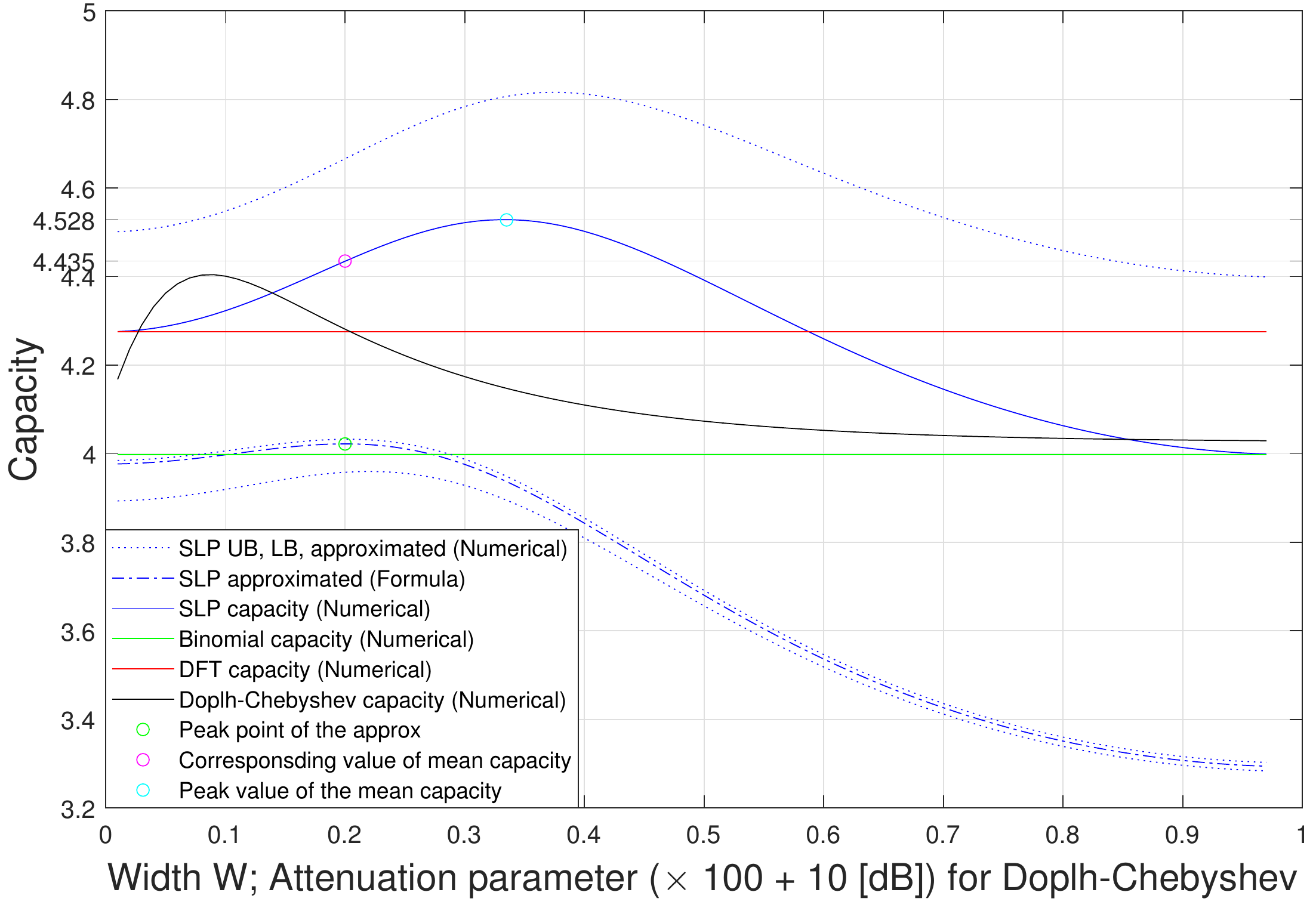}
\caption{Comparison between capacities of synthesizers.}
\label{fig:sinr}
\end{figure}

Fig.\ref{fig:sinr} shows capacity of the new synthesizer (referred SLP) versus DFT, binomial, and \emph{Dolph-Chebyshev}.
A ULA is used with $M=5$, $P_s = 1$, $\sum_n P_{I,n}=0.6$, and $N_0=0.1$.
The \emph{x-axis} draws $W$, and \emph{attenuation parameter} of \emph{Dolph-Chebyshev} to search its highest peak.
The mean capacity, the upper/lower bounds (UB/LB) and the approximation are depicted versus $W$.
Space $\mathcal{S}_{s}$ is divided into 5 equal regions so that the width that maximizes BF gain for one region is $W=0.2$ (green circle), which corresponds to a point on the mean capacity curve (magenta circle) that is higher than other synthesizers, however not the global maximum (blue circle).
All curves are depicted numerically and an analytical curve of the approximation is provided for comparison.

\begin{figure}[!t]
\centering
\includegraphics[width=3.34in]{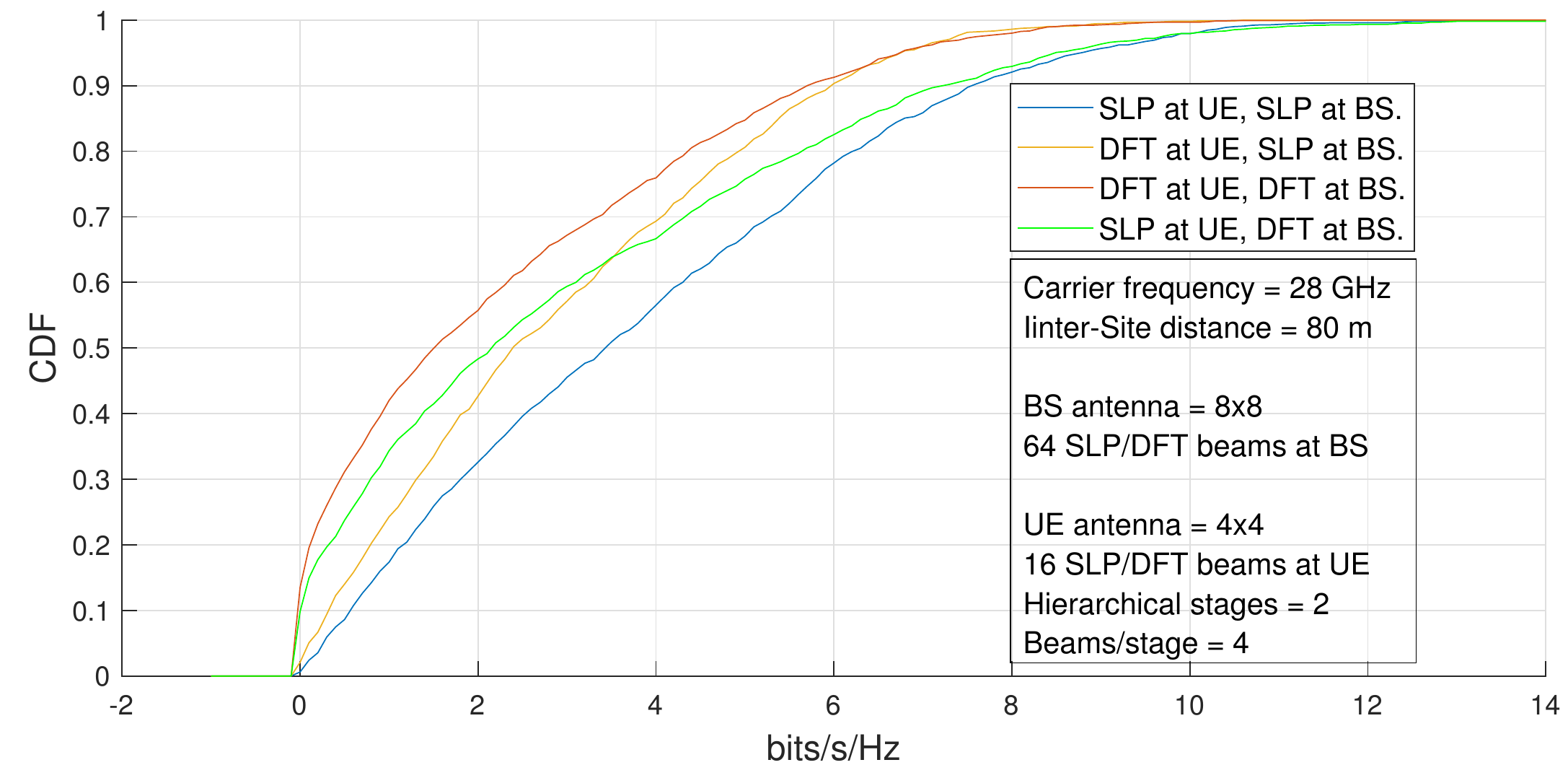}
\caption{Comparison between the beam vs DFT at BS/UE}
\label{fig:HDvsDFT}
\end{figure}

Fig.\ref{fig:HDvsDFT} is generated using a \emph{System Simulator} based on \cite{38.901}.
It shows the capacity \emph{cumulative density function} (CDF) of a 64-beam codebook at base station (BS) and a hierarchical two-stage codebook with 4 beams/stage at user equipment (UE).
The new beam used at UE and BS achieved up to 2 bits/s/Hz increase in 50\% outage capacity (roughly 100\% increase) compared to DFT at UE and BS, and offered up to 1.2 bits/s/Hz increase (around 50\% increase) compared to DFT at either UE or BS for this configuration.
Information about the \emph{System Simulator} and more simulation results in different scenarios are presented in \cite[pp.40-58]{8975347}.
\section{Conclusion and research extensions}\label{Sed:Conclusion}
The approach of designing a beam that is optimized to cover an angular region instead of a discrete direction is employed.
Thus, a new synthesizer is proposed that concentrates the maximum possible beamforming (BF) gain inside a target region for a given number of antenna elements and consequently minimizes BF gain outside.
This increases the expected BF gain, that amplifies the desired signal and mitigates interfering signals.
The tuneable width of the synthesizer is used to select the optimum width that achieves the highest SINR for a given region.
Although the optimization runs across the approximation, the corresponding value on the mean capacity still higher than those of other common synthesizers.
The properties of the beam are discussed, including the steering method which does not violate the optimization criteria.
The proposed analysis suggests the introduced synthesizer performs better than the commonly used ones (compared to DFT, Binomial and \emph{Dolph-Chebyshev}) particularly in the presence of interference.
Simulation results verified the same finding and showed that a codebook based on the new synthesizer at UE and/or BS achieves higher SINR and data rates compared to the DFT codebook.

In this work, two approximating methods are used, the capacity approximation and the phase domain.
The former is controlled as its behaviour is studied; the latter is not.
To avoid working in the phase domain, as mentioned in point \ref{numerical_integration_angular_domain} the weighting vectors can be generated in the angular domain by numerically evaluating the definite integral in \eqref{nested_trigonometric_integral} instead of \eqref{eq:IntegrationPhase}.
In this case \emph{ generalized prolate spheroidal functions} \cite{6773515} is the solution instead of DPSWF.
Extension of the work to a uniform planar array (UPA) can be found in \cite{8975347}.
\appendix
\section{Proof of lemma 1}\label{Positive definiteness}
Let $\mathbf{u} \in \mathbb{R}^M$ is an $M\times 1$ colomn vector, and $\mathbf{u} \neq 0$.
\begin{eqnarray}
\abs{\mathbf{u}^T \mathbf{a}(s)}^2 					& \geq & 0		\nonumber\\
\mathbf{u}^T \mathbf{a}(s) \mathbf{a}(s)^H \mathbf{u} 		& \geq & 0		\nonumber\\
\mathbf{u}^T \mathbf{A}(s) \mathbf{u} 				& \geq & 0		\label{eq:QuadFormGEQ0}
\end{eqnarray}
The polynomial in \eqref{eq:QuadFormGEQ0} is non-negative so its integral (the signed area between the curve and the \emph{x-axis}) is non-negative.
\begin{eqnarray}
\int_{-W}^{W} \mathbf{u}^T \mathbf{A}(s) \mathbf{u}\ ds & \geq & \int_{-W}^{W} 0\ ds		\nonumber\\
\mathbf{u}^T \int_{-W}^{W}\mathbf{A}(s)\ ds\ \mathbf{u} & \geq & 0					\nonumber\\
\mathbf{u}^T \mathbf{A}(W) \mathbf{u} & \geq & 0								\label{eq:IntQuadFormGEQ0}
\end{eqnarray}
The equality occurs if the bounds of the integral converges to one point causing the signed area converges to zero, so
\begin{equation}
\mathbf{u}^T \mathbf{A}(W) \mathbf{u} > 0 \ \ \ \ \ ,for\ W\neq 0					\label{eq:IntQuadFormG0}
\end{equation}
Equations \eqref{eq:IntQuadFormGEQ0} and \eqref{eq:IntQuadFormG0} are the quadratic forms of $\mathbf{A}(W)$ showing that the matrix is \emph{positive semi-definite} in the range $1\geq W \geq 0$, and \emph{positive definite} in the range $1\geq W > 0$.
\section{Proof of lemma 2}\label{Appendix:Lemma2}
For a function $g(x) = \sum_{a=0}^{M-1} z_a e^{-aj\beta \pi x}$, where coefficients $z_a$ are complex
\begin{eqnarray}
\int^1_{-1} [g(x)]^2 dx & = & \int^1_{-1}\abs{  \sum_{a=0}^{M-1} z_a e^{-aj\beta \pi x} }^2 dx						 		\nonumber	\\
		& = & \int^1_{-1} \sum_{a=0}^{M-1} 	z_a e^{-aj\beta \pi x} \sum_{b=0}^{M-1}z_b^* e^{bj\beta \pi x} dx					\nonumber	\\
		& = & \int^1_{-1} \sum_{a=0}^{M-1} z_a \sum_{b=0}^{M-1}	z_b^* e^{(b-a)j\beta \pi x} dx							\nonumber
\end{eqnarray}
for $a=b\implies e^{(b-a)j\beta \pi x} = 1$
\begin{equation}
\begin{split}
\int^1_{-1} [g(x)]^2 dx = \int^1_{-1} \sum_{a=0}^{M-1} z_a \big( z_{b}^*|_{b=a} + \sum_{\substack{b=0\\b\neq a}}^{M-1} z_b^* e^{(b-a)j\beta \pi x} \big) dx	\\
				 = \int^1_{-1} \big( \sum_{a=0}^{M-1} z_a z_a^* + \sum_{a=0}^{M-1} \sum_{\substack{b=0\\b\neq a}}^{M-1}z_a z_b^* e^{(b-a)j\beta \pi x} \big) dx	
\end{split}	\nonumber
\end{equation}
where $e^{\pm n j\beta \pi x}=\cos(n \beta \pi x)\pm j\sin(n \beta \pi x)$ and $n = 1, 2,\dots, M-1$.
 For symmetric integral bounds (i.e., $x=\pm1$) the imaginary part integrates to zero so
\begin{equation}
\int^1_{-1} [g(x)]^2 dx = 2\norm{\mathbf{z}}_2^2 + 2\sum_{a=0}^{M-1} \sum_{\substack{b=0\\ b\neq a}}^{M-1} z_a z_b^* \sinc((b-a)\beta \pi) 	\nonumber
\end{equation}
A special case is for $\beta = 1$, then $\sinc(\pm n\pi)=0$ and
\begin{equation}\label{eq:B2}
\int^1_{-1} [g(x)]^2 dx = 2\norm{\mathbf{z}}_2^2
\end{equation}
which is \emph{Parseval’s theorem}.
For arbitrary $\beta$, if the bounds are $x= \pm 1/\beta$, then
\begin{equation}\label{eq:B3}
\int^{\frac{1}{\beta}}_{\frac{-1}{\beta}} [g(x)]^2 dx = \frac{2}{\beta}\norm{\mathbf{z}}_2^2
\end{equation}
\section*{Declaration of competing interest}
The author declares that he has no known competing financial interests or personal relationships that could have appeared to influence the work reported in this paper.
\section*{Acknowledgment}
Many thanks to my colleagues at Huawei Lund R\&D Center, Dr. Dzevdan Kapetanovic for supervising my MSc's thesis, and Dr. Hu Sha for his helpful tips.
Finally, ultimate thanks to \scalebox{1.2}{$\mathcal{A}$\it{llah}} almighty.

\bibliographystyle{elsarticle-num-names}

\bibliography{bibliography}

\bio{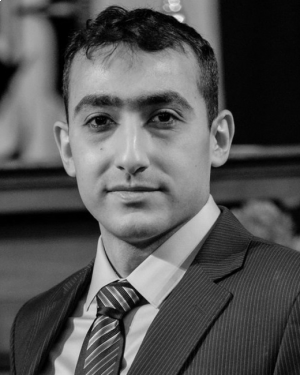}
Hesham G. SHARKAS was born in Egypt in 1985.
He received B.S. degree in electrical engineering, electronics and communications from Zagazig University, Sharkia, Egypt, in 2008 and M.S. degree in communication systems from Lund University, Skåne, Sweden, in 2019.\\
From 2008 to 2020, he was a Telecom Engineer with Systel Telecom, Cairo, Egypt.
From 2023 he conducts is PhD candidate at the department of laser measurement and navigation systems, Saint Petersburg electrotechnical university ``LETI'', Saint Petersburg, Russia.
His current research interest includes signal processing, wireless communications and photonics.
\endbio

\end{document}